\documentclass[a4paper,12pt]{article}

\newlength{\lead}
\usepackage{leading}
\leading{\lead}

\usepackage{calc}
\newlength{\abz}
\newlength{\measure}
\usepackage[a4paper, textwidth=1.15\measure, 
bindingoffset=18pt,
bottom=2.5cm,
headsep=2\lead, footskip=2\lead]{geometry}
\usepackage[utf8]{inputenc}
\usepackage[english]{babel}
\usepackage{graphicx}
\usepackage{setspace} 
\usepackage[nolist,printonlyused,smaller]{acronym} 

\usepackage[babel]{csquotes}

\usepackage{amsmath}
\usepackage{amssymb}
\usepackage{amsfonts}
\usepackage{mathrsfs} 
\usepackage{mathtools} 
\usepackage{bm} 
\usepackage{dsfont} 
\usepackage{hyperref}
\usepackage{tikz}
\usetikzlibrary{arrows,decorations.pathreplacing}
\usepackage{empheq} 
\usepackage{extarrows}
\usepackage[figuresright]{rotating}
\usepackage{slashed}

\numberwithin{equation}{section}

\newcommand{\ita}[1]{\textit{#1}}
\newcommand{\mrm}[1]{\mathrm{#1}}
\newcommand{\lef}{\left}
\newcommand{\ri}{\right}
\newcommand{\e}{\mathrm{e}}
\newcommand{\I}{\mathrm{i}}

\newcommand{\foh}{\frac{1}{2}}
\newcommand{\Tr}{\operatorname{Tr}}

\newcommand{\Det}{\operatorname{Det}}

\newcommand{\D}{\mathrm{d}}
\newcommand{\p}{\partial}
\newcommand{\sgo}{\sqrt{g}}

\newcommand{\sg}[1]{\sqrt{g(#1)}}

\newcommand{\mO}{\mathscr{O}}
\newcommand{\intD}{\int\!\mathcal{D}}
\newcommand{\mB}{\mathscr{B}}
\newcommand{\mD}{\mathcal{D}}
\newcommand{\mF}{\mathscr{F}}
\newcommand{\mE}{\mathscr{E}}

\newcommand{\mA}{\mathscr{A}}
\newcommand{\mN}{\mathscr{N}}

\newcommand{\mT}{\mathscr{T}}

\newcommand{\mM}{\mathscr{M}}

\newcommand{\mK}{\mathscr{K}}
\newcommand{\dd}{\D^d}

\newcommand{\what}[1]{\widehat{#1}}

\newcommand{\ve}{\varepsilon}

\newcommand{\bg}{\begin{equation}}
\newcommand{\eg}{\end{equation}}
\newcommand{\bgo}{\begin{equation*}}
\newcommand{\ego}{\end{equation*}}

\newcommand{\spl}[1]{\begin{split}#1\end{split}} 
\newcommand{\al}[1]{\begin{align}#1\end{align}} 
\newcommand{\alo}[1]{\begin{align*}#1\end{align*}}

\def\m{_{\mu}}
\def\n{_{\nu}}
\def\M{^{\mu}}
\def\N{^{\nu}}
\def\a{_{\alpha}}
\def\b{_{\beta}}

\def\r{_{\rho}}
\def\s{_{\sigma}}

\def\mn{_{\mu\nu}}
\def\MN{^{\mu\nu}}

\def\RS{^{\rho\sigma}}

\def\mnab{_{\mu\nu\alpha\beta}}
\def\MNAB{^{\mu\nu\alpha\beta}}

\makeatletter
\newcommand*\bigcdot{\cdot}
\newcommand*\bigcdot@[2]{\mathbin{\vcenter{\hbox{\scalebox{#2}{$\m@th#1\bullet$}}}}}
\makeatother

\DeclarePairedDelimiterX\braket[2]{\langle}{\rangle}{#1 \delimsize\vert #2} 

\newcommand{\gl}[1]{\eqref{eq:#1}}
\newcommand{\Gl}[1]{Eq.~\eqref{eq:#1}}

\newcommand{\qut}[1]{``#1''}    

\begin{document}

\begin{acronym}


\acro{qft}[QFT]{quantum field theory}

\acro{rhs}[RHS]{right-hand side}

\acro{lhs}[LHS]{left-hand side}

\acro{rs}[RS]{rigid spacetime}

\acro{sc}[SC]{selfconsistent}




\acro{uv}[UV]{ultraviolet}

\acro{ir}[IR]{infrared}

\acro{brst}[BRST]{Becchi-Rouet-Stora-Tyutin}




\end{acronym}

\begin{titlepage}

\title{
Background Independent Field Quantization\\
\mbox{}\\
with\\
\mbox{}\\
Sequences of Gravity-Coupled Approximants
}

\date{}

\author{Maximilian Becker and Martin Reuter\\[3mm]
{\small Institute of Physics, 
Johannes Gutenberg University Mainz,}\\[-0.2em]
{\small Staudingerweg 7, D--55099 Mainz, Germany}
}

\maketitle
\thispagestyle{empty}

\vspace{2mm}
\begin{abstract}

We outline, test, and apply a new scheme for nonperturbative analyses of quantized field systems in contact with dynamical gravity. While gravity is treated classically in the present paper, the approach lends itself for a generalization to full Quantum Gravity. We advocate the point of view that quantum field theories should be regularized by sequences of quasi-physical systems comprising a well defined number of the field's degrees of freedom. In dependence on this number, each system backreacts autonomously and self-consistently on the gravitational field. In this approach, the limit which removes the regularization automatically generates the physically correct spacetime geometry, i.e., the metric the quantum states of the field prefer to ``live'' in. We apply the scheme to a Gaussian scalar field on maximally symmetric spacetimes, thereby confronting it with the standard approaches. As an application, the results are used to elucidate the cosmological constant problem allegedly arising from the vacuum fluctuations of quantum matter fields. An explicit calculation shows that the problem disappears if the pertinent continuum limit is performed in the improved way advocated here. A further application concerns the thermodynamics of de Sitter space where the approach offers a natural interpretation of the micro-states that are counted by the Bekenstein-Hawking entropy.

\end{abstract}

\setstretch{1.2} 

\end{titlepage}

\newpage


\section{Introduction}

\enlargethispage{2\baselineskip}
A way of regarding a quantum field theory is as a special quantum system whose degrees of freedom are parametrized by the points of a smooth manifold. In typical applications the latter is the theoretical model of choice for representing space or spacetime. Nevertheless, when it comes to computing concrete predictions one usually discovers that such a continuum of densely packed degrees of freedom comprises, in a sense, too many of them to allow for a straightforward (or any) interpretation like in elementary quantum mechanics. The generic symptom of this overabundance are the ultraviolet divergences which relativistic quantum field theories are notorious for.\\ \indent
As a way out, a rich arsenal of tools for their \ita{regularization} and subsequent \ita{renormalization} have been devised. In many cases the first one of the two logically independent steps, regularization, is considered devoid of an immediate physical interpretation, being not more than a technical trick to render certain intermediate steps of the calculation mathematically meaningful. A classic among the many regularization schemes that are available today is the momentum space cutoff. It still possesses a certain physical flavor unlike, say, dimensional regularization or the zeta function technique. Its key ingredient is a regularization parameter which has the dimension of a mass typically, and which defines a \ita{scale} therefore: Modes of the field having momenta below this mass scale are retained by the cutoff, while the others are discarded.\\ \indent
The point to be noticed here is that the specification of a cutoff scale in proper momentum units requires a metric on the base manifold. Clearly this is no cause for concern in the familiar quantum field theories formulated on a \qut{prefabricated} and unchangeable Minkowski-space. It is a concern, however, in Quantum Gravity: when the metric itself is among the fields to be quantized one easily runs into severe difficulties or unresolvable paradoxes if one tries to work with a regularization which is metric dependent in itself \cite{Reuter:2019byg}.\\ 

\noindent\textbf{(1)} As we are going to demonstrate in the present paper, similar remarks apply already one step before full-fledged Quantum Gravity, namely within the framework of quantum field theory in curved spacetimes, if one includes the backreaction of quantum matter fields on the classical metric.\\ \indent
More generally, the purpose of this paper is as follows. We are going to investigate quantized matter fields on classical spacetime geometries whose metric is determined self-consistently by a \qut{semiclassical} Einstein equation that involves a certain effective stress tensor due to the matter degrees of freedom. To tackle this problem, we propose an approach which goes beyond earlier investigations in that it respects three basic principles which we shall introduce, explain and motivate in detail.\\ \indent 
To a large extent those principles grew out of various general lessons that were learnt within full Quantum Gravity, but have a bearing on semiclassical gravity also \cite{Cent, Ashtekar:2014kba}. They arose both in approaches to Quantum Gravity that build upon discrete structures at the fundamental level such as Loop Quantum Gravity \cite{Ehlers:1994oaa, AA-ehlers-fried, Ashtekar:2014kba} or Causal Dynamical Triangulations \cite{Ambjorn:2012jv}, as well as continuum approaches like Asymptotic Safety \cite{Reuter:2019byg,Percacci:2017fkn}.\\ \indent
The approach we are advocating, while continuum-based, could in principle detect discreteness at the physical level, should it emerge in some theory. Furthermore, in a companion paper \cite{N-2} we extend the framework presented here by also quantizing the gravitational field itself.\\ 

\noindent\textbf{(2)} The first and foremost among the three requirements is \ita{Background Independence}. The gravitational interaction has the unique property of also being in charge of furnishing the stage all physics takes place on, namely the spacetime manifold. Background Independence requires that the corresponding \qut{furniture}, the metric in particular, is obtained as the solution of some fundamental dynamical law rather than through an ad hoc selection \qut{by hand} \cite{Isham-Prima, Sei-Stam-book, Giu-BI}.\\ \indent
The second principle is an extension of Background Independence into the realm of the regularized theories, i.e., of the \qut{approximants} we deal with as long as the regulator has not yet been removed. Contrary to the examples mentioned above, we insist that the regularization should yield the approximants which are, or come close to being, realizable physical systems in their own right. More precisely, we require two properties: First, those systems possess a well defined number of degrees of freedom, and second, they are coupled to gravity and thereby respect Background Independence in the sense that their respective approximation of the spacetime metric emerges dynamically from a self-consistency condition.\\ \indent
The third principle finally is of a more technical nature and describes how to set up the approximant systems pertaining to a given field theory. The requirement is to employ what we call \ita{$N$-type cutoffs}. By definition, they amount to regularization schemes which do not involve the metric. Thus detaching cutoffs from scales allows us to take limits of the approximants that could not be considered within the standard approaches.\\

In Section \ref{sec:framework} of this paper we describe these requirements and the new framework in more detail. In the subsequent sections we shall then present a first application, which is both instructive in its own right, and can shed new light on a particularly puzzling aspect of the cosmological constant problem \cite{Weinberg:1988cp,Carroll:2000fy,Straumann:1999ia,Straumann:2002tv}, namely the gravitational field generated by the zero point oscillations of quantum fields.\\ \indent
W. Pauli is credited for the first estimate of the influence quantum vaccuum fluctuations should have on the curvature of spacetime \cite{Pauli-Calc, Straumann:1999ia}. Considering a free massless field on Minkowski space, with dispersion relation $\omega(\mathbf p)=|\mathbf p|$, he argued that the field is equivalent to a set of harmonic oscillators, and each of them should contribute its zero point energy density $\foh\omega$ to the energy of the vacuum state. Summing them up leads to an amount
\bg
\label{eq:vac}
\varrho_\mrm{vac}=\foh\int\!\frac{\D^3\mathbf{p}}{(2\pi)^3}\,|\mathbf p|
\eg
which is quartically \ac{uv} divergent and needs regularization. Installing a momentum cutoff $|\mathbf p|\leq\mathscr P$, the result is $\varrho_\mrm{vac}=c\mathscr{P}^4$ with $c$ a positive constant of order unity.\footnote{Its precise value depends on inessential implementational details.} The argument then continues by giving a numerical value to $\mathscr P$, typically taken to be the energy scale up to which the matter field theory under consideration is believed to be valid.\\ \indent
Only at this stage gravity comes into play. It is argued that like any other form of energy, $\varrho_\mrm{vac}$ should contribute to the curvature of spacetime, and that this effect can be taken care of by adding the contribution $\Delta\Lambda=(8\pi G)\varrho_\mrm{vac}=8\pi c G\mathscr{P}^4$ to the cosmological constant in Einstein's equation.\pagebreak\\ \indent
Here, then, comes the big disappointment. For every plausible scale $\mathscr P$, the curvature produced by $\Delta\Lambda$ is by far too large to be consistent with observation. Pauli himself identified $\mathscr P$ with typical energies in atomic physics; he had to conclude that the resulting curvature is so tremendous that, if correct, the universe \qut{would not even reach out to the moon} \cite{Pauli-Calc, Straumann:1999ia}. If we choose the Planck scale instead, $\mathscr P=m_\mrm{Pl}$, the calculation produces a curvature which is about $10^{120}$ times larger than the value from modern-day cosmological observations.\\ \indent
According to a variant of this reasoning, Einstein's equation contains, besides $\Delta\Lambda$, also a bare cosmological constant, $\Lambda_\mrm{b}$, whose value is then tuned in dependence on $\mathscr P$ in such a way that the sum $\Lambda_\mrm{obs}=\Lambda_\mrm{b}(\mathscr P)+\Delta\Lambda(\mathscr P)$ equals precisely the observed value. This version of the argument avoids making a false prediction (any prediction, in fact), but at the expense of an enormous naturalness problem. To achieve the desired value of $\Lambda_\mrm{obs}$, the bare quantity $\Lambda_\mrm{b}(\mathscr P)$ must be fine-tuned with a precision of $120$ digits, say.\\

One of the purposes of the present paper is to pinpoint precisely why the above reasoning must lead to a wrong answer. As it turns out employing the new and more powerful scheme no comparable \qut{cosmological constant problem} arises.\\

\noindent\textbf{(4)} The rest of this paper is organized as follows. In Section \ref{sec:framework} we outline our framework for an improved nonperturbative analysis of quantum fields in contact with dynamical gravity; in particular we explain and motivate the three basic requirements the corresponding calculations must meet. In the subsequent sections we apply those rules to a Gaussian scalar field self-consistently interacting with gravity: In Section \ref{sec:first-type} we derive a first type of approximant, in Section \ref{sec:seq-self} we explore the sequences of self-gravitating systems it gives rise to, and in Section \ref{sec:path-int} we analyze them from a path integral perspective, thereby also discovering a second type of natural approximants. While the primary application of our results is to the cosmological constant issue, Section \ref{sec:desitter} is devoted to a brief discussion of the Bekenstein-Hawking entropy of de Sitter space, and the natural interpretation of its micro-states we are led to. The final Section \ref{sec:conclusions} contains a short summary and the conclusions.

\section{The Framework: Outline and Motivation}
\label{sec:framework}

In this paper we advocate a scheme for the quantization of matter fields, and subsectors thereof, which are coupled to classical gravity.\footnote{In a companion paper \cite{N-2} we generalize the scheme by including a quantized gravitational field.} This scheme satisfies three essential requirements; in the present section, we are going to explain and motivate them, and then in the rest of the paper we implement them in various sample calculations.\\ 

The three requirements are:
\alo{
&\textbf{(R1)}\ \text{Background Independence}\\
&\textbf{(R2)}\ \text{Gravity-coupled approximants}\\
&\textbf{(R3)}\ \text{$N$-type cutoffs}
}
The requirements are not independent logically. In particular \textbf{(R2)} may be seen as an extension of \textbf{(R1)}, while \textbf{(R3)} is a tool for dealing with \textbf{(R2)}. We discuss them in turn now, focusing on the main aspects, and leaving aside inessential technical details or difficulties.

\subsection{First requirement: Background Independence}

The desideratum of Background Independence is presumably the most powerful and far reaching concept that has been taken over from Classical General Relativity and integrated into the modern approaches to Quantum Gravity \cite{Ashtekar:2014kba,Thiemann:2007pyv,Loll:2019rdj, Kie-Why}. Depending on how they cope with this challenge, the various approaches can be grouped into two classes: those which, literally, do not use background structures like a metric, and those which do employ such fields, but at a certain point fix them self-consistently, namely by invoking the fundamental dynamical laws \cite{Isham-Prima, Giu-BI}. In this paper, we develop a continuum-based approach which follows the second strategy. It enforces Background Independence indirectly by invoking the background field technique in its general form \cite{DeWitt:2003pm,Reuter:2019byg}.

\subsection{Second requirement: Gravity-coupled approximants}

The idea is to replace the notion of \ita{regularization} by sequences of certain \qut{quasi-physical} auxiliary systems describing matter; they are comprised of a well-defined, finite set of quantum degrees of freedom which couple to gravity. Referring to such systems as \ita{approximants}, we denote them symbolically by $\mathsf{App}$.\\ \indent
The total configuration of an approximant is characterized by a quantum mechanical state $\Psi_\mathbf{f}$ of the matter system, having $\mathbf f<\infty$ degrees of freedom, together with a classical metric. Symbolically, $\mathsf{App}(\mathbf f)\sim\Psi_\mathbf{f}\otimes\text{metric}$.\\ \indent
Approximants complying with the requirement \textbf{(R2)} are special in that their metric is fixed dynamically rather than by fiat. It arises as the gravitational response to the energy and momentum of the $\mathbf f$ matter degrees of freedom. They are allowed to backreact self-consistently on the geometry of the spacetime which they inhabit. Thus, by virtue of \textbf{(R2)}, Background Independence is manifest already at the regularized level.\\ \indent
Regularized quantum field theories are represented by sequences of approximants, $\mathsf{App}(\mathbf f)$, $\mathbf f\to\infty$. The removal of the regulator, corresponding in the standard case to, say, sending a lattice constant to zero, amounts to following a particular sequence for increasing $\mathbf f$. If the sequence has a limit, in an appropriate sense, we identify this limit with the \ac{qft} to be constructed, with the field in a particular state.\\ \indent 
Importantly, by this construction the (state of a) \ac{qft} arises always \ita{in combination with a self-consistently determined metric}. Symbolically,
\bg
\label{eq:S1}
\mathsf{App}(\mathbf f)\xrightarrow{\mathbf f\to\infty}\Psi_\text{\ac{qft}}\otimes\,\text{self-consistent metric} \, .
\eg
Thus, loosely speaking, possible states of the \ac{qft} for the matter sector are already \qut{born} in that particular spacetime which they like to live in. More precisely, the self-consistent metric in \gl{S1} is a solution to the semiclassical Einstein equation with an appropriate energy-momentum tensor $T\mn[\Psi_\text{\ac{qft}}]$ on its \ac{rhs}.\\

\noindent\textbf{(A) Motivations} for insisting on the requirement \textbf{(R2)} include:\\

\noindent\textbf{(A1)} We want the approximants, in the best case, to constitute physically realizable systems in their own right, or at least come close to this ideal.\footnote{As for being \qut{physically realizable} or \qut{quasi-physical} we are very liberal. What is important here is only that the drawbacks described below in connection with the counter examples are avoided.} We expect that this enhances our chances to find sequences which converge to physically interesting limits. We believe that this property is particularly important if one is forced to resort to approximate calculations of some kind, as it is always the case in practice.\\

\noindent\textbf{(A2)} As for the assumption of \ita{self-consistently} gravitating approximants subject to classical General Relativity, this is a conceptually natural requirement if one regards matter \ac{qft}s in curved spacetime as an approximation to full fledged Quantum Gravity that would additionally involve a quantized metric. In the present paper, our framework for matter \ac{qft}s is set up in such a way that it will generalize straightforwardly to full Quantum Gravity. The only difference is that for the time being quantum fluctuations of the metric, relative to the self-consistently adjusting background, are neglected.\footnote{See however \cite{N-2} for the inclusion of such metric fluctuations.} Hence, for now, the metric is obtained from the classical Einstein equation, with a quantum mechanical $T\mn[\Psi_\mathbf{f}]$, though.\footnote{Often (but somewhat confusingly) called the \qut{semiclassical Einstein equation}.}\\

\noindent\textbf{(A3)} There is a further motivation for physical gravity-coupled approximants which goes beyond using the sequences of approximants merely as a tool for regularizing a \ac{qft}. Namely, our framework is open towards the possibility that experiment tells us that Nature is actually better described by the approximant $\mathsf{App}(\mathbf{f}_\mrm{obs})$ for some finite, observationally determined $\mathbf{f}_\mrm{obs}$, rather than by $\mathsf{App}(\mathbf f)$ in the limit $\mathbf f\to\infty$. In this hypothetical case, observational facts would suggest to abandon the original plan of removing the regulator fully. The quantization of a classical field would then result in a quantum system with, rigorously, only finitely many degrees of freedom.\\ \indent
While this might sound like a quite exotic possibility, there is a natural and simple scenario which realizes it: Assume that the quantum gravitational dynamics is such that spacetime acquires physical, i.e., observable discreteness properties at some microscopic scale. It is a question of general significance then whether, and possibly how, this effect can be discovered by a continuum based theoretical framework.\footnote{See Section 1.5 of \cite{Reuter:2019byg} for a discussion of this point.} One possibility is the scenario above: If spacetime is roughly similar to a certain discrete point set with matter fields on it, it will, loosely speaking, have the appearance of an \qut{incompletely quantized} classical field. Later on we shall encounter an explicit example where this is indeed what happens.\\

\noindent\textbf{(B) Counter examples} are perhaps the best way to characterize, and to further motivate, approximants that comply with \textbf{(R2)}:\\

\noindent\textbf{(B1)} Many regularization schemes that we employ routinely because they are convenient technically fail to generate quasi-physical approximants in the sense of \textbf{(R2)}. A typical example is \ita{dimensional regularization}. Clearly it is impossible to interpret a regularized theory at a generic value of $\ve\equiv 4-d$ as a quantum system with a defined number $\mathbf f$ of degrees of freedom which an experimentalist could build. Similar remarks apply to other schemes based upon analytical continuation such as \ita{zeta function regularization} \cite{Hawking:1976ja, Dowker:1975tf} which, too, is unacceptable by the requirement \textbf{(R2)}.\\

\noindent\textbf{(B2)} Furthermore, \textbf{(R2)} rejects all schemes which are designed so as to make certain (usually power-law) divergences invisible, with the justification that they anyhow would be absorbed into bare parameters whose values do not matter for all practical purposes. Zeta function regularization is an example again. It is notorious for \qut{identities} like
\bg
\label{eq:S10}
\sum_{n=1}^\infty n\overset{\text{zeta fct.}}{\equiv}-\frac{1}{12}
\eg
which exhibit a finite part on top of a numerically much more important divergent one -- which it suppresses however. By way of comparison, a scheme that does comply with \textbf{(R2)} is a cutoff regularization that would deal with the divergent sum $\sum_n n$ simply by stating that
\bg
\label{eq:S11}
\sum_{n=1}^N n=\foh N^2\lef\{1+O\!\lef(\frac{1}{N}\ri)\ri\} \, ,
\eg
without expressing any prejudice about the ultimate fate of the divergence which arises when $N\to\infty$ at a later stage.\\ \indent
As a rule, acceptable regularization schemes must treat bare parameters as \ita{potentially physical}. Hence contributions like the $N^2$-term above must be retained, and their physical impact along the sequence of approximants must be taken into account carefully.\\

\begin{figure}[ht]
	\centering
  \includegraphics[width=0.9\textwidth]{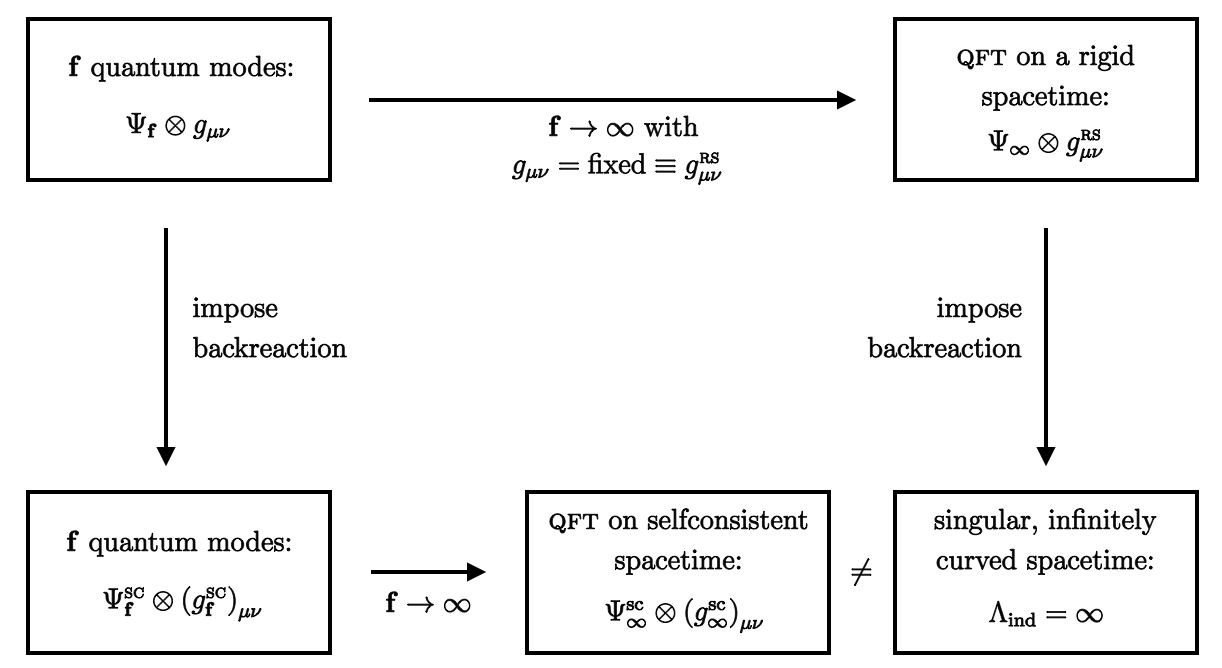}
	\caption{Inclusion of the gravitational backreaction does not commute with the limit $\mathbf f\to\infty$.}
	\label{fig:D}
\end{figure}

\noindent\textbf{(B3)} Finally we turn to the perhaps most important issue, the question of why gravity should be included into the physical description of the approximants.\\ \indent 
Figure \ref{fig:D} contrasts field quantization via the sequences $\{\mathsf{App}(\textbf f)\}$ with the standard approach. The horizontal arrows of this highly schematic diagram represent taking the limit $\mathbf f\to\infty$, i.e., the transition from a finite system to a \ac{qft}, while the vertical arrows symbolize the inclusion of the gravitational backreaction into the matter system by solving a semiclassical Einstein equation.
\\ \indent
The upper left box in Figure \ref{fig:D} stands for a generic regularized precursor of the \ac{qft}, which does not in general comply with \textbf{(R2)}. It consists of quasi-physical systems in states $\Psi_\mathbf{f}\otimes g\mn$,  with $\mathbf f<\infty$, but the systems live in a spacetime with an arbitrary fixed metric $g\mn$, which is unrelated to $\mathbf f$ and the quantum states $\Psi_\mathbf{f}$ a priori.\\ \indent
Now there exist two paths in the diagram which one can take in order to remove the regulator. Symbolically speaking they are, respectively,
\al{
&\textbf{(i)}\ \text{first down, then horizontal}\label{eq:S15}\\
&\textbf{(ii)}\ \text{first horizontal, then down}\label{eq:S16}
}

\noindent\textbf{The path (i)} is the one advocated in the present paper: \ita{For each approximant separately} we solve the respective Einstein equation. It contains $T\mn[\Psi_\mathbf{f}]$ which is self-consistently coupled to the Schr{\"o}dinger equation for $\Psi_\mathbf{f}$; the latter involves the metric which is a solution to the former. In this manner we arrive at a sequence of configurations describing the total system, $\Psi_\mathbf{f}^\text{\ac{sc}}\otimes\lef(g_\mathbf{f}^\text{\ac{sc}}\ri)\mn$, $\mathbf f=0,1,2,\cdots$.
Thereafter, as the second step, we let $\mathbf f\to\infty$ and (ideally) are led to a limiting state $\Psi_\infty^\text{\ac{sc}}\otimes\lef(g_\infty^\text{\ac{sc}}\ri)\mn$ which can be interpreted as pertaining to a full fledged \ac{qft} living on a spacetime with metric $\lef(g_\infty^\text{\ac{sc}}\ri)\mn$. It is selected dynamically by self-consistency, in accord with Background Independence.\\

\noindent\textbf{The path (ii)} is the familiar one; it does not respect \textbf{(R2)} and the price one pays for that is the fatal cosmological constant problem in calculations like Pauli's. At first one takes the field theory limit leading to infinitely many degrees of freedom, and thereby one does not yet worry about gravity at all. The construction of the \ac{qft} is performed on a \ac{rs} whose metric $g\mn^\text{\ac{rs}}$ is chosen \ita{in a completely arbitrary way}, often highly symmetric, typically flat, so as to ease the calculational difficulties. In this \ac{qft}, one then computes the vacuum energy density, or induced cosmological constant $\Lambda_\mrm{ind}$, and finds that it is formally infinite in absence of a cutoff.\footnote{Or it is finite, but still way too large if the \ac{qft} is regarded an \ita{effective} one which is valid below a certain plausible \ac{uv} scale only.}\\ \indent
\enlargethispage{\baselineskip}
Only now, in the second stage one asks about the backreaction of the vacuum energy on the metric. The answer is very simple then: If $\Lambda_\mrm{ind}=\infty$, the curvature is infinite and spacetime is degenerate. And even in an effective field theory setting, the predicted curvature is too large, by factors like $10^{120}$ or so. The lower right box in Figure \ref{fig:D} indicates this singular, or at least phenomenologically unacceptable, state of the combined \ac{qft}-gravity system.\footnote{Unless stated otherwise, bare parameters are always kept fixed when $N\to\infty$ in this paper. If one allows them to depend on $N$, the $(\Lambda_\mrm{ind}=\infty)$-problem can be traded for the naturalness problem of an infinite finetuning.}\\

The main goal of the present paper is to demonstrate by various sample calculations that the state obtained along track \textbf{(i)}, $\Psi_\infty^\text{\ac{sc}}\otimes\lef(g_\infty^\text{\ac{sc}}\ri)\mn$, can be extremely different from the singular one of \textbf{(ii)}. In fact, choosing track \textbf{(i)} the kind of cosmological constant problem one encounters in the Pauli-type approaches does not occur.\\ \indent
In essence, the requirement \textbf{(R2)} wants us to perform any \ac{qft} calculation in curved spacetime by proceeding in analogy with track \textbf{(i)} of this example. This guarantees that Background Independence is respected not only by the final theory, but already at the level of its approximants. As a result, essential dynamical effects come within reach of practically feasible calculations.

\subsection{Third requirement: $N$-type cutoffs}

Given a classical field theory, the question is how to actually manufacture sequences of approximants by means of which we can hope to find interesting limits that would qualify as a \ac{qft} coupled to gravity. We propose to define the approximants by imposing a cutoff which is of \qut{$N$-type}, a concept which we outline next. It may be seen as a generalization of the ``finite-mode regularization'' that had been used before in a non-gravitational context~\cite{FM1,FM2,FM3}.\\ \indent 
The expression $N$-type cutoff derives from the fact that in typical examples the regularization parameter is a positive integer, $N\in\mathds{N}$, but other cases will occur as well.\\ \indent 
Using an $N$-type cutoff, the number of degrees of freedom $\mathbf f\equiv\mathbf f(N)$ becomes a function of the cutoff parameter $N$. In the case $N\in\mathds N$ regularized \ac{qft}s are thus represented by sequences $\{\mathsf{App}(N)\ ,\ N=0,1,2,\cdots\}$.

\subsubsection{$N$-cutoffs: definition}

Let us consider the general problem of giving a meaning to a formal functional integral of the type
\bg
\label{eq:S20}
Z=\int_\mathcal{F}\!\mD(C)\,\e^{-S[C]} \, .
\eg
The integration is over fields $C(x)$ that \qut{live} on some differentiable manifold, $\mM$, and belong to a certain space of functions, $\mathcal F$. We assume that the space $\mathcal F$ is the span of the basis
\bg
\label{eq:S21}
\mathscr B=\lef\{ w_\alpha(\,{\bigcdot}\,)\ \big|\ \alpha\in I\ri\}
\eg
where $I$ is an appropriate index set. So we can expand the integration variable,
\bg
\label{eq:S22}
C(x)=\sum_{\alpha\in I}c_\alpha w_\alpha(x) \, ,
\eg
and reexpress the measure $\mD(C)$ in terms of the expansion coefficients $c_\alpha$. In typical examples the result will be proportional to $\prod_{\alpha\in I}\D c_\alpha$ then.\\ 

Nothing has been gained so far; \gl{S20} is still an ill-defined combination of infinitely many integrals. To regularize it, we introduce a one parameter family of subsets $\mathscr B_N\subset\mathscr B$ labeled by a dimensionless number $N\in\mathds{N}$ (or $N\in [0,\infty)\equiv\mathds{R}^+$).\footnote{Our general description focuses on $N\in\mathds N$ mostly since our later examples will belong to this case. The most important features of an $N$-cutoff carry over to the case $N\in\mathds{R}^+$ straightforwardly, however. Note also that as such field systems having only a finite number of modes, $N\in\mathds N$, are nothing exotic, of course. This is the usual situation in condensed matter physics~\cite{Datta}, or lattice field theory~\cite{Gernot}, for example.} This family of subsets is required to satisfy
\bg
\label{eq:S23}
\mathscr B_0=\varnothing\quad,\quad\mathscr B_\infty=\mathscr B\quad,\quad\text{and}\quad N_2>N_1\Rightarrow\mathscr B_{N_2}\supset\mathscr B_{N_1} \, .
\eg
Thus, for increasing values of $N$, a continually growing portion of the basis elements in $\mathscr B$ get included into the set $\mathscr B_N$.\\ \indent
Eligible sequences of subsets
\bg
\label{eq:S24}
\mB_0\subset\mB_1\subset\mB_2\subset\cdots\subset\mB_N\subset\dots\subset\mB
\eg
should be described entirely in terms of the indices in the set $I$. For each $N\in\mathds{N}$ we introduce a subset of indices $I_N\subset I$ such that
\bg
\label{eq:S25}
\mB_N=\lef\{ w_\alpha(\,{\bigcdot}\,)\ \big|\ \alpha\in I_N\ri\} \, .
\eg
Then the specification of a sequence like \gl{S24} boils down to a long \qut{bit string} of yes-no decisions: For any given element of $\mB$, say $w\b$, which is uniquely identified by its label $\beta\in I$, we must specify whether $\beta\in I_N$ or $\beta\not\in I_N$, for all $N=0,1,2,\cdots,\infty$. We stress that no length or momentum scales are involved in the specification of such sequences $\{\mB_N\}_{N=0,\cdots,\infty}$.\\ \indent
The rationale behind the $\mB_N$s is that integration over their linear span, denoted by $\mathcal F_N\subset\mathcal F$, should lead to a well defined regularized precursor of the functional integral, viz.
\bg
\label{eq:S26}
Z_N\equiv\int_{\mathcal F_N}\!\mD(C)\,\e^{-S[C]} \, .
\eg
In the case $N\in\mathds N$, \gl{S26} involves only finitely many integrals. (If $N\in\mathds{R}^+$ we assume that $\mathcal F_N$ is defined \qut{sufficiently small} to make \gl{S26} mathematically meaningful.) We regard $Z_N$ as a partition function which describes an approximant with a sufficiently small, or even finite number of degrees of freedom, $\mathbf f\equiv\mathbf f(N)$. They are realized by $\mathbf f$ selected modes of a field on $\mM$.\\ \indent
Replacing the full space of functions $\mathcal F$ successively by the chain of subspaces $\{\mathcal F_N\}_{N=0,\cdots,\infty}$ is what we will, very broadly, refer to as a regularization by means of a \ita{cutoff of the $N$-type}.

\subsubsection{$N$-cutoffs: properties}

The most important property which an $N$-cutoff must possess (while many others don't) is the following: \ita{Assigning a particular value to $N$ does not imply a momentum or length scale} separating the modes retained in $\mB_N$ from those discarded.\\ \indent
\enlargethispage{\baselineskip}
\noindent\textbf{(1)} The benefit which we get from the complicated-looking definition above is that \ita{$N$-cutoffs can be formulated without the need of a metric}. In fact, $\mB$ and $\mB_N$ are certain bases of functions on a given differentiable manifold $\mM$. Hence technically speaking they belong to the manifold's differentiable structure, and to define the cutoff no metric structure is required on $\mM$.\\ \indent
Above we therefore insisted that the $\mB_N$s are defined in terms of binary decisions operating on the index set $I$. This highlights the importance of adequately labeling the basis functions: The enumeration of the $\mB$-elements, i.e., the map $I\to\mathcal F$, $\alpha\mapsto w\a(\,{\bigcdot}\,)$, the selection of the $\mB_N$-elements and their enumeration by $I_N\to\mathcal F$, $\alpha\mapsto w\a(\,{\bigcdot}\,)$, must not involve a metric in any way.\\

\noindent\textbf{(2)} Before discussing practical incarnations of $N$-cutoffs let us clarify another point. It is essential for our purposes to keep the steps of \ita{regularization} and \ita{renormalization} strictly separated. By definition, $N$-type cutoffs are really not more than a regularization, and they must not implement any renormalization conditions implicitly.\footnote{Recall the counter example of the zeta function regularization which unavoidably fixes finite parts in its own specific way.}\\ \indent
At least for standard field theories, and in absence of gravity, there is a well known procedure that can be followed when one tries to define a continuum limit\footnote{We follow the common practice and refer to the limit where the cutoff is removed generally as the \qut{continuum limit} also when the regularization is not by a lattice.} of regularized path integrals like \gl{S26}. Namely, while the regulator is removed, one tries to change the bare parameters (masses, couplings, etc.) which are implicit in $S[C]$ in precisely such a manner that the $Z_N$s do indeed converge to some limit. If this is (im-)possible, the theory in question is called (non-)renormalizable \cite{Glimm-Jaffe}.\\ \indent
Clearly the same can be done on the basis of the above sequences of approximants. However, as it turns out it is also of interest to take the approximants at finite $N$ seriously in their own right, and to compare them as physical systems. They are in different states $\Psi_N$, but refer to \ita{the same values of the bare parameters}. The examples worked out in the present paper will be of this second kind.

\subsubsection{Eigenbases of metric dependent operators}

Next we discuss a special type of $N$-cutoffs. While particularly convenient from the technical point of view, superficially it might seem that they are inconsistent with \textbf{(R3)}. It is therefore important to see that this is not the case actually.\\ \indent
Let us now explicitly assume that $\mM\equiv(\mM,g\mn)$ is a Riemannian manifold, and let us allow the path integral to depend manifestly on the metric $g\mn$ as a background:
\bg
\label{eq:S30}
Z[g]=\intD(C;g)\,\e^{-S[C;g]} \, .
\eg
Furthermore, let $\mathscr K\equiv\mathscr K[g]$ be a self-adjoint positive operator depending on the metric, the prime example being the (negative of the) Laplace-Beltrami operator, $\mK=-\Box_g\equiv-g\MN D\m D\n$. We would like the basis $\mB\equiv\mB[g]=\{w\a[g](\,{\bigcdot}\,)\ |\ \alpha\in I\}$ to be an eigenbasis of $\mK[g]$ now, implying that in general the basis functions $w\a(\,{\bigcdot}\,)\equiv w\a[g](\,{\bigcdot}\,)$ and their eigenvalues $\lambda\a[g]$ will have a parametric dependence on $g\mn$:
\bg
\label{eq:S31}
\mK[g]\, w\a[g](x)=\lambda\a[g]\, w\a[g](x)\quad,\quad\alpha\in I \, .
\eg \indent
From here on we proceed in the usual way in order to install an $N$-cutoff in \gl{S30}. We declare, for all $N$, which indices $\alpha\in I$ are in $I_N$, so that setting
\bg
\label{eq:S32}
\mB_N[g]=\lef\{w\a[g](\,{\bigcdot}\,)\ \big|\ \alpha\in I_N\ri\}
\eg
yields a sequence $\{\mB_N\}$ which obeys the general rules \gl{S32}, and we can define the approximants by
\bg
\label{eq:S33}
Z_N[g]\equiv\!\!\int\displaylimits_{\mrm{span}\,\mB_N[g]}\!\!\!\!\!\!\mD(C;g)\,\e^{-S[C;g]} \, .
\eg\indent
It is important to understand why this regularization still amounts to an $N$-cutoff in accordance with \textbf{(R3)}: While the \ita{elements} of the set $\mB_N[g]$ do indeed depend on the metric, the crucial property is that \ita{the index set $I_N$ is metric independent}. This is why we repeatedly emphasized that (in the discrete case) the sets $I_N$ should be the result of nothing but \qut{binary decisions} applied to the elements of $I$; as such they do not define a proper length or mass scale.\\

To summarize: An $N$-cutoff, even when applied to a formal functional integral with an explicit metric dependence, is such that its specification in terms of $\{I_N\}$ does not require, and does not involve a metric.\\ \indent
A simple example illustrates this point. Let $(\mM,g)$ be the round $2$-sphere with radius $r$. Then $\mK=-\Box_g$ has the well known spectrum $l(l+1)/r^2$, $l=0,1,2,\cdots$, which is linked to the metric via the value of $r$. Now we can specify an $N$-cutoff by decreeing that, for example, $\mB_N$ contains all spherical harmonics with $l\leq N$ and none having $l>N$, i.e., $\mB_N=\{Y_{lm}\ |\ (l,m)\in I_N\}$ where $I_N=\{(l,m)\ |\ l=0,1,2,\cdots,N\ ;\ m=-l,\cdots,+l\}$. The $\mB_N$s are completely fixed upon specifying the rule according to which index pairs from the full $I\equiv I_\infty$ are allocated to $I_N$, and this rule has nothing to do with the continuous parameter $r$, i.e., with the metric which we put on the sphere.

\subsubsection{Continuous spectra}

In order to demonstrate that $N$-cutoffs are not restricted to discrete spectra, we briefly consider the example of a  foliated cosmological spacetime equipped with the correspondingly adapted spatially flat Robertson-Walker metric
\bg
\label{eq:S40}
\D s^2=-\D t^2+a(t)^2\,\delta_{ij}\D x^i\D x^j \, .
\eg
\enlargethispage{\baselineskip}
Here coordinate differences $(\Delta x^i)\equiv\Delta\mathbf x$, while often referred to as \ita{comoving distances}, are not proper distances. Products $a(t)\Delta\mathbf x$ instead are \qut{proper} with respect to the metric \gl{S40}. Let us take $\mK$ to be the spatial part of the corresponding Laplace-Beltrami operator,
\bg
\label{eq:S41}
\mK=-\frac{1}{a(t)^2}\delta^{ij}\p_i\p_j \, .
\eg
Its eigenfunctions are plane waves clearly. A subtlety arises however when it comes to labeling them, since later on $N$-cutoffs are defined in terms of the pertinent index set.\pagebreak\\ \indent
From this perspective, a \qut{good} labeling amounts to writing
\bg
\label{eq:S42}
w_\mathbf{q}(\mathbf x)=\exp(\I\mathbf q\cdot\mathbf x)\quad,\quad\mathbf q\in\mathds{R}^3 \, ,
\eg
in terms of the \ita{coordinate} momentum $\mathbf q$; like $\mathbf x\equiv(x^1,x^2,x^3)$, it is dimensionless, not \qut{proper}, and in fact unrelated to any metric. It qualifies as a continuous version of the generic (multi-)indices $\alpha$ by means of which the $\mB_N$-elements are selected. Hence a perfectly legitimate $N$-cutoff would, for example, be specified by
\bg
\label{eq:S43}
\mB_N=\lef\{w_\mathbf{q}(\,{\bigcdot}\,)\ \big|\ \mathbf{q}\in I_N\ri\}\quad\text{with}\quad I_N=\lef\{\mathbf q\in\mathds{R}^3\ \big|\ |\mathbf q|\leq N\ri\} \, .
\eg
In this case $N\in\mathds{R}^+$, so the sequence of approximants is labeled in a continuous fashion now.\\ \indent
However, there are also frequently used ways of labeling the eigenfunctions which are \qut{bad} from the perspective of \textbf{(R3)}. In fact, the eigenvalue of \gl{S42} is given by
\bg
\label{eq:S44}
\lambda_\mathbf{q}=\lef(\frac{\mathbf{q}}{a(t)}\ri)^2\equiv\mathbf{p}^2 \, .
\eg
This motivates using the proper momentum $\mathbf p\equiv\mathbf q/a(t)$ in order to distinguish the eigenmodes of $\mK$, rewriting \gl{S42} as
\bg
\label{eq:S45}
W_\mathbf{p}(\mathbf{x})=\exp\!\Big(\I a(t)\mathbf p\cdot\mathbf x\Big)\quad,\quad\mathbf p\in\mathds{R}^3 \, .
\eg
When working with the mode functions \gl{S45} it would seem natural to impose a cutoff condition on the proper momentum, like $|\mathbf p|\leq\mathscr P$, say. But as we discuss next, this would violate \textbf{(R3)}.

\subsubsection{$N$-cutoffs vs. $\mathscr P$-cutoffs}

We close the discussion of the $N$-type cutoffs by exhibiting a class of counter examples which we collectively refer to as cutoffs of \qut{$\mathscr P$-type}. They fail to satisfy the requirement \textbf{(R3)} because of a \qut{mistake} that can be pinned down quite precisely. Later on we shall then see that this \qut{mistake} has a significant impact on the cosmological constant issue.\\ \indent
We return to the spectral problem \gl{S31} and, in order to simplify the argument, assume that all eigenvalues are non-degenerate. Hence, for $g\mn$ fixed, there exists a one-to-one map $\alpha\mapsto\lambda\a[g]$ which relates eigenvalues $\lambda$ and labels $\alpha$. It is a common practice to solve this relationship for the label, obtaining $\alpha=\alpha[g](\lambda)$, and to \ita{use the eigenvalues in order to enumerate the eigenfunctions}. The basis writes then
\bg
\label{eq:S50}
\mB[g]=\lef\{W_\lambda[g](\,{\bigcdot}\,)\ \big|\ \lambda\in\mrm{spec}(\mK)\ri\}
\eg
with the reparametrized mode functions
\bg
\label{eq:S51}
W_\lambda[g](x)\equiv w\a[g](x)\bigg|_{\alpha=\alpha[g](\lambda)} \, .
\eg \indent
Now we come to the delicate point: Being presented with the basis in the form of \gl{S50} it is tempting to construct regularizations by applying selection criteria to the new label $\lambda$, in the same way as with $\alpha$ above. Of course, the first example that comes to mind is a sequence $\{\mB_\mathscr{P}\}_{\mathscr{P}\in\mathds{R}^+}$ obtained by restricting the eigenvalues to lie below a fixed scale $\mathscr{P}^2$:
\bg
\label{eq:S52}
\mB_\mathscr{P}=\lef\{ W_\lambda(\,{\bigcdot}\,)\ \big|\ \lambda\leq\mathscr{P}^2\ri\} \, .
\eg \indent
Obviously the familiar momentum space cutoff that is used abundantly in field theory on flat space is precisely of this sort, with $\mathscr P\equiv\Lambda$ in the traditional notation. The background metric, $g\mn=\delta\mn$~or~$\eta\mn$ usually, is fixed once and for all in this case.\\ \indent
Nonetheless, the subsets $\mB_\mathscr{P}$ in \gl{S52} do not define a cutoff in accord with \textbf{(R3)}, i.e., no $N$-cutoff. The reason is obvious: Due to the substitution $\alpha\to\alpha[g](\lambda)$ the \ita{enumeration} of the basis functions has become explicitly metric dependent. As we explained above, an \qut{adiabatic} $g\mn$-dependence of $w\a[g](x)$ is perfectly acceptable -- as long as the labeling by the $\alpha$s does not involve the metric. The new mode functions $W_\lambda[g](x)$ spoil this property. As a result, regularizations like \gl{S50} which are defined in terms of their \qut{index} $\lambda$ are not $N$-type cutoffs.\\ \indent
Note also that $\lambda$ and $\mathscr P$, unlike $\alpha$ and $N$, are not dimensionless: They have canonical mass dimensions $[\lambda]=2$ and $[\mathscr P]=1$, respectively.\\

An example of such a forbidden metric-related labeling is \gl{S45} in the cosmological example above. The proper momentum $\mathbf p$ has a magnitude determined by the eigenvalue, $|\mathbf p|^2=\lambda$, and a direction specified by a unit vector $\mathbf n\equiv\mathbf p/|\mathbf p|$, which serves as a dimensionless degeneracy index here. Hence \gl{S45} is equivalent to writing
\bg
\label{eq:S55}
W_{\lambda,\mathbf n}(\mathbf x)=\exp\lef(\I a(t)\sqrt{\lambda}\,\mathbf n\cdot\mathbf x\ri) \, .
\eg
This simple example also makes it clear that our argument generalizes trivially to spectra with degenerate eigenvalues if $\lambda$ is combined with appropriate degeneracy indices.\\ \indent
In the following we shall refer to regularizations of the form \gl{S52} collectively as \ita{cutoffs of the $\mathscr P$-type}.

\section{A First Type of Approximants}
\label{sec:first-type}

In the rest of this paper we perform an explicit investigation which implements all three requirements \textbf{(R1)}, \textbf{(R2)} and \textbf{(R3)} simultaneously. As a model system, we consider scalar particles which couple to a classical gravitational field but do not interact among themselves. The present section covers the steps leading to the finite approximant systems.

\subsection{The classical field}

Our theoretical laboratory is a free scalar field $A(x)$ which lives on a classical $d$ dimensional spacetime $\mM$ that carries an externally prescribed Euclidean metric $g\mn$. We assume that $\mM$ is compact and has no boundary, and that the dynamics of $A(x)$ is governed by the matter action
\bg
\label{eq:P1}
\spl{
S[A;g]&=\foh\int_\mM\!\dd x\sgo\,\lef\{g\MN D\m A\, D\n A+M^2 A^2+\xi R A^2\ri\}\\
&=\foh\int_\mM\!\dd x\sgo\,A\,\mK A
}
\eg
with a selfadjoint kinetic operator ($\Box_g\equiv D^2\equiv g\MN D\m D\n$)
\bg
\label{eq:P2}
\mK=-\Box_g+M^2+\xi R(x) \, .
\eg
Stationarity of $S[A;g]$ with respect to $A$ implies the equation of motion
\bg
\label{eq:P3}
\lef[-\Box_g+M^2+\xi R(x)\ri] A(x)=0 \, ,
\eg
while its functional derivative with respect to the metric gives rise to the Euclidean stress tensor:
\bg
\label{eq:P4}
T\MN[A;g](x)=-\frac{2}{\sgo}\frac{\delta}{\delta g\mn(x)}S[A;g] \, .
\eg
For arbitrary parameters $M$ and $\xi$ we have 
\bg
\label{eq:P5}
\spl{
T\MN[A;g]=D\M A\, D\N A&-\foh\, g\MN\lef[ g\RS D\r A\, D\s A+M^2A^2+\xi R A^2\ri]\\
&+\xi\lef[R\MN-D\M D\N+g\MN D^2\ri] A^2 \, .
}
\eg
Evidently the Rosenfeld type stress tensor \gl{P4} is symmetric. If evaluated on a solution to the equation of motion, $A\equiv A^\mrm{sol}$, it is also well known to be conserved and, under certain conditions, traceless:
\al{
D\m T\MN[A^\mrm{sol};g]&=0 \, ,\label{eq:P6}\\
g\mn T\MN[A^\mrm{sol};g]&=0\quad\text{if}\quad\xi=\xi_c(d)\quad\text{and}\quad M=0 \, .\label{eq:P7}
}
Here we abbreviated $\xi_c(d)\equiv\frac{d-2}{4(d-1)}$.

\subsection{The quantum system at finite $N$}

Next, we employ the above field theory as a classical inspiration in setting up a quantum mechanical system with finitely many degrees of freedom. Concretely, we identify those degrees of freedom with the $N$ lowest eigenvalues of the kinetic operator.\\

\noindent\textbf{(1) The spectral problem.} Given the metric $g\mn$, we construct the operator \gl{P2} and consider its eigenvalue problem on $\mM$:
\bg
\label{eq:P10}
\mK u_{n,m}(x)=\mF_n\, u_{n,m}(x) \, .
\eg
The discrete eigenvalues $\mF_n$ are enumerated by an integer $n=0,1,2,3,\cdots$ which labels them in ascending order: $\mF_0<\mF_1<\mF_2<\cdots$. Allowing for a $D_n$-fold degeneracy of $\mF_n$, the eigenfunctions $u_{n,m}(x)$ carry an additional degeneracy index, or multi-index, $m$. By analogy with the generalized spherical harmonics \cite{Rubin:1983be,Rubin:1984tc} we may think of the indices $n$ and $m$ as a kind of angular momentum and magnetic quantum number, respectively.\\ \indent
The eigenfunctions $\{ u_{n,m}\ |\ n\geq0\ ;\ m=1,2,\cdots,D_n\}\equiv\mB$ form a complete set of scalar functions on $\mM$. They can be orthonormalized with respect to the inner product on $L^2(\mM)$ supplied by the metric,
\bg
\label{eq:P10-1}
\int\!\dd x\sgo\,u_{n,m}^*(x)\,u_{\bar n,\bar m}(x)=\delta_{n\bar n}\delta_{m\bar m}
\eg
so that the corresponding completeness relation reads
\bg
\label{eq:P10-2}
\sum_{n,m} u_{n,m}(x)\,u_{n,m}^*(y)=\frac{\delta(x-y)}{\sg{y}} \, .
\eg

\noindent\textbf{(2) Definition of the approximants.} Given the basis $\mB$, we define the quantum mechanical system $\mathsf{App}(N)$ by truncating the set $\mathscr B$ at the level $n=N<\infty$, retaining only the eigenfunctions of $\mK$ with eigenvalues $\mF_n\leq\mF_N$. Instead of arbitrary fields $A(x)$, we consider only those that can be expanded in the truncated basis $\mB_N\equiv\{u_{n,m}\ |\ n=0,1,\cdots,N\ ;\ m=1,\cdots,D_n\}$, i.e.,
\bg
\label{eq:P11}
A(x)=\sum_{n=0}^N\sum_{m=1}^{D_n}\alpha_{n,m}\,u_{n,m}(x) \, .
\eg
The degrees of freedom of the quantum system are represented then by the coefficients $\{\alpha_{n,m}\ |\ 0\leq n\leq N\ ,\ m=1,\cdots,D_n\}$. Their total number equals
\bg
\label{eq:P12}
\sum_{n\leq N}D_n\equiv\mathbf f(N) \, .
\eg
In a path integral treatment (which will be the topic of Section~\ref{sec:path-int} below) the functional integral over $A(x)$ reduces to an integration over the finitely many coefficients $\alpha_{n,m}$s then. In view of the trucated expansion \gl{P11} it is suggestive to visualize the spacetime of $\mathsf{App}(N)$ as a fuzzy sphere \cite{fuzzy}.\\

\noindent\textbf{(3) The 2-point function.} Ordinarily, the key ingredient of a free Euclidean field theory is the 2-point correlation function $\langle\what A(x)\what A(y)\rangle^g\equiv G(x,y)$.\footnote{Depending on the context the caret notation ($\what A$, etc.) indicates operators or integration variables under a functional integral. Furthermore, the notation $\langle\,{\bigcdot}\,{\bigcdot}\,{\bigcdot}\,\rangle^g$ emphasizes that all expectation values must be regarded functionals of the metric on $\mM$.} By Wick's theorem it determines all higher $n$-point functions, and it satisfies
\bg
\label{eq:P15}
\mK G(x,y)=\frac{\delta(x-y)}{\sg{x}}
\eg
with suitable boundary conditions being specified. In the case $N\to\infty$, the completeness relation \gl{P10-2} gives rise to a formal solution of \gl{P15}, namely\footnote{Here and in the following it is understood that if $\mK$ has a zero mode it is separated off in the usual way. We do not indicate this notationally.}
\bg
\label{eq:P16}
G(x,y)=\sum_{n=0}^\infty\sum_{m=1}^{D_n}\frac{u_{n,m}(x)\,u_{n,m}^*(y)}{\mF_n} \, .
\eg\indent
For the time being, we define the quantum theory of the finite field system $\mathsf{App}(N)$ by the correspondingly truncated version of \Gl{P15}, namely
\bg
\label{eq:P17}
\mK G_N(x,y)=\frac{1}{\sg{x}}\,\delta_N(x,y)
\eg
whereby now both the regularized correlator,
\bg
\label{eq:P18}
G_N(x,y)\equiv\sum_{n=0}^N\sum_{m=1}^{D_n}\frac{u_{n,m}(x)\,u_{n,m}^*(y)}{\mF_n}
\eg
and the modified delta function,
\bg
\label{eq:P19}
\delta_N(x,y)\equiv\sg{x}\,\sum_{n=0}^N\sum_{m=1}^{D_n}u_{n,m}(x)\,u_{n,m}^*(y) \, ,
\eg
are constructed in terms of functions from the truncated set $\mB_N$ only.\\

\noindent\textbf{(4) Bilinear observables.} In the following we are mostly interested in observables that are bilinear in $\what A(x)$. With the regularized 2-point function $G_N(x,y)=\langle\what A(x)\what A(y)\rangle_N^g$ at hand it is in principle straightforward to calculate their expectation values in the state $G_N$ corresponds to. Thanks to the \ac{uv} cutoff, $G_N(x,y)$ is non-singular in the limit $y\to x$, and so all those observables have well defined expectation values. They can be expressed by finite sums over the functions $u_{n,m}$.\\ \indent
The local monomial $D\m\what A(x)D\M\what A(x)$, for instance, leads to
\al{
\lef\langle(D\m\what A)^2(x)\ri\rangle&=\lim_{y\to x}\lef\langle\p\m^x
\what A(x)\,\p\M_y\what A(y)\ri\rangle_N^g\notag\\
&=\lim_{y\to x}\p\m^x\p\M_y G_N(x,y) \label{eq:P20}\\
&=\lim_{y\to x}\p\m^x\p\M_y\sum_{n\leq N,m}\frac{1}{\mF_n}u_{n,m}(x)u_{n,m}^*(y)\notag\\
&=\sum_{n\leq N,m}\frac{1}{\mF_n}D\m u_{n,m}(x)\,D\M u_{n,m}^*(x)\notag
}
which is perfectly finite as long as $N<\infty$.\pagebreak\\ \indent
A particularly interesting integrated monomial is $\int\!\dd x\sgo\,\what A(x)\mK\what A(x)$. Its expectation value counts the number of degrees of freedom which the quantum mechanical system $\mathsf{App}(N)$ possesses:
\al{
\lef\langle\int\!\dd x\sgo\,\what A(x)\mK\what A(x)\ri\rangle_N^g&=\int\!\dd x\sg{x}\,\lim_{y\to x}\mK_x\lef\langle \what A(x)\what A(y)\ri\rangle_N^g\notag\\
&=\int\!\dd x\sg{x}\,\lim_{y\to x}\sum_{n\leq N,m}\frac{1}{\mF_n}\mK_x u_{n,m}(x)u_{n,m}^*(y)\notag\\
&=\sum_{n\leq N,m}\int\!\dd x\sg{x}\,u_{n,m}(x)u_{n,m}^*(x)\label{eq:P21}\\
&=\sum_{n\leq N} D_n\notag\\
&=\mathbf f(N) \, .\notag
}
Here we also exploited the eigenvalue equation and the normalization condition satisfied by the mode functions.\\

\noindent\textbf{(5) Trace of the stress tensor.} The most important bilinear operator is the energy-momentum tensor of the field modes that inhabit $\mM$. Thanks to the $N$-cutoff, the operator $T\MN[\what A;g]$ which we obtain from the classical expression \gl{P5} by letting $A(x)\to \what A(x)$ is not plagued by any operator product singularities.\\ \indent
Nevertheless, and this is important to be kept in mind, there is always an ambiguity with regard to the \qut{correct} energy-momentum tensor of $\mathsf{App}(N)$. As always in quantum mechanics, the classical expression for an observable is at best an \qut{inspiration} when guessing the quantum operator. After all, the two can differ by any number of $O(\hbar)$ terms that disappear in the classical limit.\\ \indent
Therefore, if we now \ita{declare} that $T\MN[\what A;g]$ as given by \gl{P5} with $A\to\what A$ is the correct energy-momentum tensor of our quantum mechanical system, this amounts to a choice over and above the decision for an $N$ cutoff.\\ \indent
\enlargethispage{2\baselineskip}
We shall need in particular the operator which represents the trace of $T\MN$. It writes, without using the field equations,
\bg
\label{eq:P25}
\spl{
T\M\m[\what A;g]=&\,\foh\lef[2-d+4(d-1)\xi\ri] D\m\what A\,D\M\what A\\
&+2(d-1)\xi\ \what A\, D\m D\M\what A\\
&+\foh(2-d)\xi R\ \what A^2-\foh d\, M^2\,{\what A}^2 \, .
}
\eg
The expectation value of \gl{P25} is easily obtained by the same steps as above:
\al{
\lef\langle T\M\m[\what A;g](x)\ri\rangle_N^g=\foh\sum_{n\leq N,m}\frac{1}{\mF_n}\bigg\{&\lef[2-d+4(d-1)\xi\ri]D\m u_{n,m}^*(x) D\M u_{n,m}(x)\notag\\
&+4(d-1)\xi\,u_{n,m}^*(x) D\m D\M u_{n,m}(x)\label{eq:P26}\\
&+\lef[(2-d)\xi\,R(x)-d\, M^2\ri] u_{n,m}^*(x) u_{n,m}(x)\bigg\} \, .\notag
}
Up to this point, $g\mn$ is an arbitrary externally prescribed metric. While it is usually difficult to solve the spectral problem of $\mK$ in a concrete case, it is clear that in principle \gl{P26} and its un-traced analogue provide us with welldefined finite expectation values for any choice of $g\mn$.

\subsection{Backreaction on the metric}
\label{subsec:back-metric}

Next we promote the spacetime metric $g\mn$ to a dynamical, yet still classical, quantity which responds to the energy and momentum carried by the quantum fluctuations of the finite field system $\mathsf{App}(N)$. We assume this system to be in its ground state. Classically this means that $A=0$ and hence $T\mn=0$ everywhere on $\mM$. Instead, quantum mechanically, the vacuum fluctuations of the $\mathbf f$ degrees of freedom contribute to the energy and momentum in the universe which determine the metric on $\mM$.\\ \indent 
We assume that the metric is governed by the semiclassical Einstein equation
\bg
\label{eq:P30}
R\mn-\foh g\mn R+\Lambda_\mrm{b}\,g\mn=8\pi G\lef\langle T\mn[\what A;g]\ri\rangle_N^g \, 
\eg
where $\Lambda_\mrm{b}$ is a bare cosmological constant. Importantly, the \ac{rhs} of the equation \gl{P30} involves the same metric $g\mn$ as its \ac{lhs}, both explicitly via the operator $T\mn[\what A;g]$, and implicitly throgh the expectation value. This is what makes solutions to \gl{P30} \ita{self-consistent}.\\ \indent 
We denote such self-consistently determined metrics by $(g_N^\text{\ac{sc}})\mn$ in the following.
The question we shall be particularly interested in concerns the dependence of the solutions on the parameter $N$, and thus on the number $\mathbf f(N)$ of degrees of freedom living on $\mM$.\\ \indent
In full generality the semiclassical Einstein equation \gl{P30} represents an extremly hard problem; in principle the expectation value involved must be computed as an explicit functional of $g\mn$. It is given by sums over eigenfunctions like in \gl{P26}. Evaluating them requires first of all solving the spectral problem of $\mK\equiv\mK[g]$ for \qut{all} metrics $g\mn$.\\

\noindent\textbf{(1) Maximally symmetric spacetimes.} To make some progress here, we restrict the space of metrics to those of maximally symmetric Riemannian spaces of positive curvature, i.e., spheres $S^d(L)$. They come with only a single free parameter, namely the radius $L$, a Euclidean version of the Hubble length. We write the metric on $S^d(L)$ in the form
\bg
\label{eq:P31}
g\mn(x)=L^2\,\gamma\mn(x)
\eg
where $\gamma\mn$ is the dimensionless metric on the unit $d$-sphere.\footnote{Our conventions concerning the assignment of canonical mass dimensions are such that $[x\M]=[\p\m]=[\D^d x]=[\delta(x)]=0$, while $[g\MN]=-[g\mn]=[R]=[\Box_g]=[\mF_n]=+2$, $[g]\equiv[\det(g\mn)]=-2d$, $[\gamma\mn]=0$ and $[A]=\foh(d-2)$, $[g^{1/4}\D A]=-1$, $[\mu]=-[L]=[\mathscr P]=+1$.} Thus, the determination of self-consistent background geometries of the type $\mM=S^d(L)$ boils down to finding the $N$-dependence of $L\equiv L^\text{\ac{sc}}(N)$.\\ \indent 
The curvature scalar on spheres is a constant,
\bgo
R=R(L)\equiv\frac{d(d-1)}{L^2} \, .
\ego
Therefore the eigenfunctions $u_{n,m}$ of $\mK=-\Box_g+M^2+\xi R$ coincide with those of the scalar Laplacian $\Box_g=D\m D\M$, and the eigenvalues of $\mathscr K$ are
\bg
\label{eq:P35}
\mF_n=\mE_n+M^2+\xi R
\eg
with $\{\mE_n\}$ denoting the spectrum of $-\Box_g$:
\bg
\label{eq:P36}
-\Box_g u_{n,m}(x)=\mE_n\, u_{n,m}(x) \, .
\eg
The eigenvalues $\mE_n$ and their multiplicities $D_n$ are well known \cite{Rubin:1983be,Rubin:1984tc}:
\al{
\mE_n&=\frac{n(n+d-1)}{L^2}\equiv\mE_n(L)\quad,\quad n=0,1,2,\cdots \, ,\label{eq:P37}\\
D_n&=\frac{(2n+d-1)(n+d-2)!}{n!(d-1)!} \, .\label{eq:P38}
}\indent
On the sphere, $\Box_g$ has a zero mode, the constant function appearing at $n=0$. We exclude this mode from the degrees of freedom belonging to $\mathsf{App}(N)$. So, to be precise, their total number equals
\bg
\label{eq:P39}
\mathbf f(N)=\sum_{n=1}^N D_n \, .
\eg
This, and all similar sums appearing below start at $n=1$. However, for the present analysis the precise treatment of the low lying modes plays no role; the relevant regime will always be dominated by $n\gg 1$.\\

\noindent\textbf{(2) The effective Einstein equation.} Thanks to the maximum symmetry of the background geometry we have $\langle T\mn\rangle_N^g\propto g\mn$, and so it suffices to analyze the traced, and now, $x$-independent Einstein equation:
\bg
\label{eq:P40}
-\foh(d-2)\,R(L)+d\,\Lambda_\mrm{b}=8\pi G\lef\langle T\m\M[\what A;g]\ri\rangle_N^g \, .
\eg
Moreover, no information is lost when we integrate \gl{P40} over spacetime:
\bg
\label{eq:P41}
\lef\{-\foh(d-2)R(L)+ d\,\Lambda_\mrm{b}\ri\}\mrm{Vol}[S^d(L)]=8\pi G\int\!\dd x\sgo\,\lef\langle T\m\M[\what A;g]\ri\rangle_N^g \, .
\eg
The virtue of the latter integration is that, upon inserting our earlier result \gl{P26} into \gl{P41}, it allows us to perform an integration by parts on the $D\m u^* D\M u$-terms, and then to simplify the entire sum by exploiting \gl{P36} and the orthonormality of the $u_{n,m}$s.\\ \indent 
This brings us to the main result of this section, namely the following condition for self-consistency:
\bg
\label{eq:P42}
-\foh(d-2)\,R(L)+d\,\Lambda_\mrm{b}=8\pi G\,\frac{\Theta_N(L)}{\mrm{Vol}[S^d(L)]} \, .
\eg
Its main building block is the dimensionless and manifestly finite mode sum representing the integrated trace of the stress tensor:
\bg
\label{eq:P43}
\spl{
\Theta_N(L)&\equiv\int\!\dd x\sgo\,\lef\langle T\m\M[\what A;g]\ri\rangle_N^g\\
&=-\sum_{n=1}^N D_n\lef[\foh(d-2)+\frac{M^2}{\mE_n(L)+\xi R(L)+M^2}\ri] \, .
}
\eg
Furthermore, the volume in \Gl{P42} is given by
\bg
\label{eq:P44}
\mrm{Vol}[S^d(L)]=\frac{2\pi^{(d+1)/2}}{\Gamma((d+1)/2)}L^d \, .
\eg
Several remarks are in order at this point.\\

\noindent\textbf{(3) Limiting cases.} For vanishing and very large (infinite) mass we obtain, respectively,
\al{
\Theta_N(L)\big|_{M=0}&=-\foh(d-2)\,\mathbf f(N)\label{eq:P48}\\
\Theta_N(L)\big|_{M\to\infty}&=-\foh d\,\mathbf f(N) \, .\label{eq:P49}
}
Note that the limit $M\to\infty$ is performed at fixed, finite $N$.\\

\noindent\textbf{(4) Background dependent counterpart.} It should be emphasized that the respective $L$-dependencies on the \ac{lhs} and \ac{rhs} of the reduced Einstein equation \gl{P42} have a very different origin: The one on the \ac{lhs} stems from the metric to be found, i.e., the one in the Einstein tensor. On the \ac{rhs} instead, in $\Theta_N(L)/\mrm{Vol}[S^d(L)]$, the radius $L\equiv L_\text{\ac{rs}}$ refers to a logically different metric of a certain rigid spacetime (\qut{\ac{rs}}) in which energy and momentum of the vacuum fluctuations are computed. Hence, in a traditional \ita{background dependent} calculation, \Gl{P42} would appear replaced by
\bg
\label{eq:P45}
-\foh(d-2)R(L)+d\,\Lambda_\mrm{b}=8\pi G\frac{\Theta_N(L_\text{\ac{rs}})}{\mrm{Vol}[S^d(L_\text{\ac{rs}})]} \, .
\eg
Herein $L_\text{\ac{rs}}$ is an absolute constant, fixed by hand once and for all. In the literature a popular choice of such a rigid spacetime is flat space $(L_\text{\ac{rs}}\to\infty)$ since the corresponding mode functions are technically easiest to deal with.\\ \indent
However, in the rest of this paper we shall have many opportunities to see that solving \gl{P45} at fixed $L_\text{\ac{rs}}$ can result in sequences $L=L(N)$ which are very different from those derived in a Background Independent fashion, namely by first setting $L=L_\text{\ac{rs}}$ and solving Einstein's equation thereafter. We believe that the $L_\text{\ac{rs}}$-based sequences convey a physically wrong picture of what happens in the limit $N\to\infty$.\\

\noindent\textbf{(5) Four dimensions.} In the following sections we shall analyze the condition for self-consistent maximally symmetric backgrounds in detail and explore the corresponding sequences
\bg
\label{eq:P49}
N\mapsto L^\text{\ac{sc}}(N;\xi,M,G,\Lambda_\mrm{b})
\eg
in dependence on the various bare parameters. We specialize for $d=4$ dimensions then, so that the self-consistency condition becomes
\bg
\label{eq:P50}
-R(L)+4\Lambda_\mrm{b}=\frac{3G}{\pi}\frac{\Theta_N(L)}{L^4}
\eg
with $R(L)=12/L^2$, and
\bg
\label{eq:P51}
-\Theta_N(L)=\sum_{n=1}^N D_n\lef[1+\frac{(ML)^2}{n(n+3)+12\xi+(ML)^2}\ri] \, .
\eg
In $4$ dimensions, $\mE_n=n(n+3)/L^2$, and the multiplicities of the eigenvalues are governed by the cubic polynomial
\bg
\label{eq:P52}
D_n=\frac{1}{6}(2n+3)(n+2)(n+1) \, .
\eg
One easily proves by mathematical induction that the sum \gl{P39} evaluates to
\bg
\label{eq:P53}
\mathbf f(N)=\sum_{n=1}^N D_n=\frac{1}{12}[N^4+8N^3+23N^2+28N] \, .
\eg
This is the total number of degrees of freedom inhabiting our spherical spacetimes in the four dimensional case.\\

\noindent\textbf{(6) The fuzzy $\mathbf{S^4}$}. On the $4$-sphere, the eigenfunctions $u_{n,m}\equiv Y_{n{l_1}{l_2}m}$ are labeled by four integer quantum numbers. Besides $n=0,1,2,\cdots$ which determines the eigenvalue, there is a triple of integers $(l_1,l_2,m)$ satisfying $n\geq l_1\geq l_2\geq |m|$. They play the role of the degeneracy multi-index $m$ now. The $S^4$-harmonics depend on four angular coordinates and have the general structure
\bg
\label{eq:P54}
Y_{n{l_1}{l_2}m}(\zeta,\eta,\vartheta,\varphi)\propto\,{}_4\bar P_n^{l_1}(\zeta)\,{}_3\bar P_{l_1}^{l_2}(\eta)\,{}_2\bar P_{l_2}^m(\vartheta)\,\frac{1}{\sqrt{2\pi}}\e^{\I m\varphi}
\eg
where the ${}_i\bar P_k^j$ denote generalized associated Legendre functions, see \cite{Higuchi:1986wu} for a detailed account.\\ \indent
Recalling the construction of the approximants, an interesting property is the \qut{resolving power} of the basis functions \gl{P54} when one restricts $n\leq N$ in the series expansions. It is not difficult to show that functions $A(x)\equiv A(\zeta,\eta,\vartheta,\varphi)$ represented by such truncated series can display nontrivial structures down to angular separations of approximately \cite{Reuter:2005bb}
\bg
\label{eq:P55}
\Delta\alpha\approx\frac{\pi}{N} \, .
\eg
The minimum proper distance that can be resolved by the truncated basis is about $\Delta\ell\approx\pi L/N$ then. In this sense, the spacetimes of the approximants are reminiscent of \qut{fuzzy spheres} \cite{fuzzy}.

\section{Sequences of Self-Gravitating Quantum Systems}
\label{sec:seq-self}

Next, we apply the above semiclassical Einstein equation in order to search for sequences of well behaved self-gravitating quantum systems enumerated by the cutoff quantum number $N$. The member labeled \qut{$N$}, $\mathsf{App}(N)$, consists of $\mathbf f(N)$ quantized field degrees of freedom. They inhabit a spacetime whose metric, $(g_N^\text{\ac{sc}})\mn$, they decide about in a self-determined, democratic way.

\subsection{Massless scalar field}

We begin by considering a massless scalar field in $4$ dimensions. Setting $M=0$ in \gl{P51} has the very welcome consequence that the sum that is to be evaluated boils down to nothing but the counting function \gl{P53}:
\bg
\label{eq:P60}
-\Theta_N(L)=\sum_{n=1}^N D_n=\mathbf f(N)=\frac{N^4}{12}\lef\{1+O\lef(\frac{1}{N}\ri)\ri\} \, .
\eg
We see that upon setting $M=0$ the integrated trace $\Theta_N$ automatically becomes independent of $L$ and $\xi$ also. It is given by a pure number, namely the number of degrees of freedom the quantum system possesses. (The relative minus sign in $\Theta_N=-\mathbf f$ is a consequence of our Euclidean conventions for the stress tensor.) Hence the only $L$-dependence on the \ac{rhs} of the self-consistency condition \gl{P50} is due to the volume factor $\propto L^4$:
\bg
\label{eq:P61}
\frac{12}{L^2}-4\Lambda_\mrm{b}=\frac{3G}{\pi}\,\frac{\mathbf f(N)}{L^4} \, .
\eg

\noindent\textbf{(1) The classical initial point.} If we set $N=0$ there are no quantum mechanical degrees of freedom, $\mathbf f(0)=0$, and provided $\Lambda_\mrm{b}>0$, \Gl{P61} yields $L=\sqrt{\Lambda_\mrm{b}/3}$, i.e., the radius of the well known $S^4$ solution to Einstein's equation in vacuo.\\

\noindent\textbf{(2) Exactness.} Incidentally, the modified Einstein equation \gl{P61} can be reexpressed succinctly in terms of the curvature scalar as
\bg
\label{eq:P62}
R-4\Lambda_\mrm{b}=\frac{G}{48\pi}\mathbf f(N)\, R^2 \, .
\eg
The $R^2$ term on its \ac{rhs} might be reminiscent of the derivative expansions that often are calculated on the basis of the asymptotic heat kernel series. It must be stressed however that \gl{P62} actually enjoys a much more reliable status: For spherical spacetimes, the \ac{rhs} of \Gl{P62} is an exact, non-perturbative result and not merely a term in an asymptotic series. Its derivation does not involve any expansion in a small coupling or in the number of derivatives. For $M=0$ and in $4$ dimensions, curvature powers both higher and lower than $R^2$ are strictly absent.

\subsection{The background dependent calculation}

We begin the discussion of the self-consistency condition \gl{P61} by solving its background dependent couterpart. As we mentioned in Subsection \ref{subsec:back-metric}, in its $\langle T\m\M\rangle$-term it has $L$ replaced by a rigid scale $L_\text{\ac{rs}}$ independent of $L$, the radius representing the dynamical metric in the symmetry reduced case. As a consequnce, the $\langle T\m\M\rangle$-term can be treated as a contribution to the cosmological constant, yielding:
\bg
\frac{3}{L^2}=\Lambda_\mrm{tot}(N)\label{eq:P63}
\eg
with the modified cosmological constant
\bg
\Lambda_\mrm{tot}(N)\equiv\Lambda_\mrm{b}+\frac{3 G}{4\pi}\,\frac{\mathbf f(N)}{L_\text{\ac{rs}}^4} \, . \label{eq:P64}
\eg
\noindent\textbf{(1)} Now we let $N\to\infty$ with $L_\text{\ac{rs}}$ and $\Lambda_\mrm{b}$ fixed. Then the total cosmological constant behaves as $\Lambda_\mrm{tot}(N)\propto N^4\nearrow+\infty$ for sufficiently large $N$. This forces the dynamical radius to approach zero,
\bg
\label{eq:P65}
L(N)=\lef(\frac{1}{3}\Lambda_\mrm{tot}(N)\ri)^{-1/2}\propto\frac{1}{N^2}\to 0 \, ,
\eg
so that the curvature scalar of the solution grows unboundedly, $R(N)=12/L(N)^2\propto N^2\to\infty$.\\ \indent
What we encounter here is an incarnation of the cosmological constant problem as it arises from the Pauli-style calculations. They sum up the vacuum energies of a certain number of modes propagating \ita{on a fixed spacetime}, and thereafter insert the resulting energy density \ita{in one package} into Einstein's equation as part of the cosmological constant. Then either $\Lambda_\mrm{tot}$ becomes unacceptably large for any physically plausible choice of the cutoff scale, or the bare parameter $\Lambda_\mrm{b}$ must be finetuned with unnatural precision.\\ 

\noindent\textbf{(2)} In the background dependent calculation, one of the roles played by the rigid metric on $\mM=S^4(L_\text{\ac{rs}})$ consists in relating the dimensionless cutoff $N$ to a dimensionful one. The eigenvalue $\mE_N$ of $-\Box$, on this background geometry, is analogous to the dimensionful cutoff scale (traditionally denoted $\Lambda^2$) at which an ordinary momentum cutoff on flat space would become operative. In the case at hand, the pure number $N$ gives rise to the \ac{uv} cutoff scale $\mathscr P$ according to 
\bg
\label{eq:P66}
\mathscr{P}^2=\mE_N(L_\text{\ac{rs}})=\frac{N(N+3)}{L_\text{\ac{rs}}^2} \, .
\eg
Note that $\mathscr P$ and $\mE_N$ have canonical mass dimensions $+1$ and $+2$, respectively.\footnote{In the \ac{qft} literature, $\mathscr P$ is denoted as $\Lambda$ usually. We do not use this notation to avoid confusion with the cosmological constant.} Indeed, \gl{P66} is an example of a \qut{$\mathscr P$-type} cutoff.\\ \indent
It sounds like a quite trivial remark -- actually it is not, as we shall see -- that \ita{$\mathscr P$ is a monotonically increasing function of $N$,} and hence $N\to\infty$ is tantamount to $\mathscr P\to\infty$.\\ 

\noindent\textbf{(3)} Note also that when reexpressed in terms of $\mathscr P$, the curvature scalar reads, for $N\gg 1$ and $\Lambda_\mrm{b}=0$, say:
\bg
\label{eq:P67}
R=4\Lambda_\mrm{tot}=\frac{G}{4\pi}\mathscr{P}^4 \, .
\eg
When presented in this fashion, the unphysical quantity $L_\text{\ac{rs}}$ drops out from the final result.\footnote{Also typical calculations on flat space lead to a result of this form when the continuous spectrum of $4$-momenta is cut off by $p\m^2<\mathscr P^2$.} This may further contribute to the -- false, as it turns out~-- impression that invoking a rigid auxiliary spacetime during the intermediate steps of the calculation is just a harmless technical convenience.

\subsection{Self-consistent approximants: The case $\Lambda_\mrm{b}=0$}
\label{subsec:lambda-null}

Let us now begin our search for sequences of approximants which satisfy the requirement of self-consistency. For various parameter choices of interest we determine their Hubble radii $L\equiv L^\text{\ac{sc}}(N)$, and more importantly, the pertinent scalar curvatures
\bg
\label{eq:P700}
R^\text{\ac{sc}}(N)=\frac{12}{L^\text{\ac{sc}}(N)^2} \, .
\eg\indent
Sometimes it is suggestive to think of the quantum mechanical term in the effective Einstein equation, \ita{at the point of consistency}, to contribute an additional piece to the cosmological constant; this shifts $\Lambda_\mrm{b}$ to a certain $\Lambda^\text{\ac{sc}}(N)$ which satisfies, by definition,
\bg
\label{eq:P701}
R^\text{\ac{sc}}(N)=4\Lambda^\text{\ac{sc}}(N) \, .
\eg
For the specific example of \Gl{P61}, the modified cosmological constant reads
\bg
\label{eq:P702}
\Lambda^\text{\ac{sc}}(N)=\Lambda_\mrm{b}+\frac{3 G}{4\pi}\,\frac{\mathbf f(N)}{L^\text{\ac{sc}}(N)^4} \, .
\eg
Of course, the relation \gl{P702} can be employed only \ita{after} having solved Einstein's equation: this is the very difference between the background dependent treatment and the Background Independent one.\\

\noindent\textbf{(1) The $\mathbf{\Lambda_\mrm{b}=0}$ solution.} Letting $N=1,2,3,\cdots$, we now populate spacetime with an increasing number of degrees of freedom and check if \Gl{P61} admits self-consistent $S^4$ solutions.\\ \indent
In this subsection we focus on the particularly interesting case of a vanishing bare cosmological constant, $\Lambda_\mrm{b}=0$. After multiplication with $L^4$, assuming $L\neq 0,\infty$, the self-consistency condition becomes very simple therefore:
\bg
\label{eq:P70}
L^2=\frac{G}{4\pi}\,\mathbf f(N) \, .
\eg
Obviously, the quantum mechanical contribution to the Einstein equation has just the correct sign so that there does indeed exist a self-consistent solution $L\equiv L^\text{\ac{sc}}(N)$ for any number of degrees of freedom:
\bg
\label{eq:P71}
\spl{
L^\text{\ac{sc}}(N)&=\lef[\frac{G}{4\pi}\,\mathbf f(N)\ri]^{1/2}\\
&=\lef(\frac{G}{48\pi}\ri)^{1/2}N^2\lef\{1+O\lef(\frac{1}{N}\ri)\ri\} \, .
}
\eg
The scalar curvature of the spacetimes found are given by
\bg
\label{eq:P72}
\spl{
R^\text{\ac{sc}}(N)&=\frac{48\pi}{G}\cdot\frac{1}{\mathbf f(N)}\\
&=\frac{576\pi}{G}\cdot\frac{1}{N^4}\lef\{1+O\lef(\frac{1}{N}\ri)\ri\} \, .
}
\eg
This sequence of self-consistent spacetimes $\{\mathsf{App}(N)\}$ displays a number of highly surprising and unusual features:\\

\noindent\textbf{(2) Inflating spheres.} When additional scalar modes are added to the quantum system, i.e., $N$ is increased, the radius $L^\text{\ac{sc}}(N)$ becomes \ita{larger}, and the curvature $R^\text{\ac{sc}}(N)$ correspondingly \ita{smaller}. In the limit $N\to\infty$, the radius of the sphere approaches infinity, and the self-consistent spacetime that supports those infinitely many field modes approaches \ita{flat space}.\\ \indent
This behavior is the exact opposite of what we had found by means of the background dependent calculation: There, adding further modes led to a \ita{smaller} radius, higher curvature, and a larger effective cosmological constant. And in the limit $N\to\infty$ the spacetime degenerated to a point even.\\

\noindent\textbf{(3) The $\mathbf{\mathscr P}$-cutoff.} To understand the origin of the unexpected result in the Background Independent case, let us look at the dimensionful companion of the pure-number cutoff $N$. In absence of any absolute metric that could be employed to turn $N$ into a dimensionful quantity, i.e., an \ita{inverse proper length}, only the actually realized, dynamically determined self-consistent metric can be used for this purpose. Since this metric depends on $N$, the $(-\Box)$-eigenvalue at the upper end of the sequence, i.e., $\mE_N\equiv\mathscr{P}^2(N)$, acquires a second, indirect dependence on $N$ now, namely via the radius:
\bg
\label{eq:P73}
\mathscr{P}(N)^2=\frac{N(N+3)}{\lef(L^\text{\ac{sc}}(N)\ri)^2}=\frac{4\pi}{G}\cdot\frac{N(N+3)}{\mathbf f(N)} \, .
\eg\indent
Now, when $N$ increases the novel factor $\mathbf f(N)$ in the denominator of \gl{P73} grows $\propto N^4$ for large $N$, and so it overrides the familiar $N(N+3)$ in the numerator. As a consequence, the relationship between the dimensionless $N$-cutoff and its dimensionful counterpart $\mathscr P(N)$ assumes a rather unusual and unexpected form in the Background Independent case; when $N\gg 1$,
\bg
\label{eq:P74}
\mathscr P(N)=\lef(\frac{48\pi}{G}\ri)^{1/2}\frac{1}{N}\lef\{1+O\lef(\frac{1}{N}\ri)\ri\} \, .
\eg
We see that in the Background Independent setting \ita{the dimensionful cutoff $\mathscr P$ is a decreasing function of the dimensionless integer $N$}.\\ \indent
This is in sharp contradistinction to the standard result $\mathscr P\approx N/L_\text{\ac{rs}}$ of \gl{P66} which one obtains in the background dependent case.\\
\noindent\textbf{(i)} Note also that the relation \gl{P74} that connects $\mathscr P$ and $N$ depends \ita{nonanalytically} on Newton's constant which hints at its nonperturbative character.\\ 
\noindent\textbf{(ii)} While at first sight the Background Independent relationship $\mathscr P\propto 1/N$ seems rather counterintuitive, it is nevertheless easy to understand its origin:\\ \indent 
Each member in the sequence of self-gravitating systems $\{\mathsf{App}(N)\}_{N\in\mathds{N}}$ comes with its own self-consistent, that is, dynamically determined metric $(g_N^\text{\ac{sc}})\mn$, here represented by the radius $L^\text{\ac{sc}}(N)$. And each member uses its own, individual metric in order to convert its number in the sequence, $N$, to a dimensionful cutoff scale which then enjoys the status of an inverse proper length with respect to this particular metric. It is clear therefore that if $(g_N^\text{\ac{sc}})\mn$ has a sufficiently strong $N$ dependence, the emergent function $\mathscr P(N)$ no longer has any reason to depend on $N$ monotonically.\\ \indent
In the case at hand we encounter the extreme situation where the $N$ dependence of the metric is so strong that increasing $N$ actually even lowers the mass scale of the \ac{uv} cutoff, $\mathscr P(N)$.\\
\noindent\textbf{(iii)} It is also remarkable that upon using \gl{P74} to eliminate $N$ in favor of $\mathscr P$, \Gl{P72} assumes the form
\bg
\label{eq:P78}
R^\text{\ac{sc}}(N)=\frac{G}{4\pi}\mathscr P(N)^4 \, .
\eg
This is exactly the same relationship between the curvature and $\mathscr P$ as in \Gl{P67} which resulted from the background dependent calculation. An yet, there remains the crucial difference that now in \gl{P78} both $R^\text{\ac{sc}}(N)\to 0$ and $\mathscr P(N)\to 0$ when $N\to\infty$.

\subsection{Self-consistent approximants: The case $\Lambda_\mrm{b}\neq 0$}

Next, we admit a nonvanishing bare cosmological constant $\Lambda_\mrm{b}$. For now we assume it independent of $N$. Provided $L\neq 0,\infty$, the self-consistency condition \gl{P61} is equivalent to a quadratic equation for $L^2$:
\bg
\label{eq:P80}
\frac{2}{3}\Lambda_\mrm{b}L^4-2L^2+\frac{G}{2\pi}\mathbf f(N)=0 \, .
\eg

\begin{figure}[ht]
	\centering
  \includegraphics[width=0.7\textwidth]{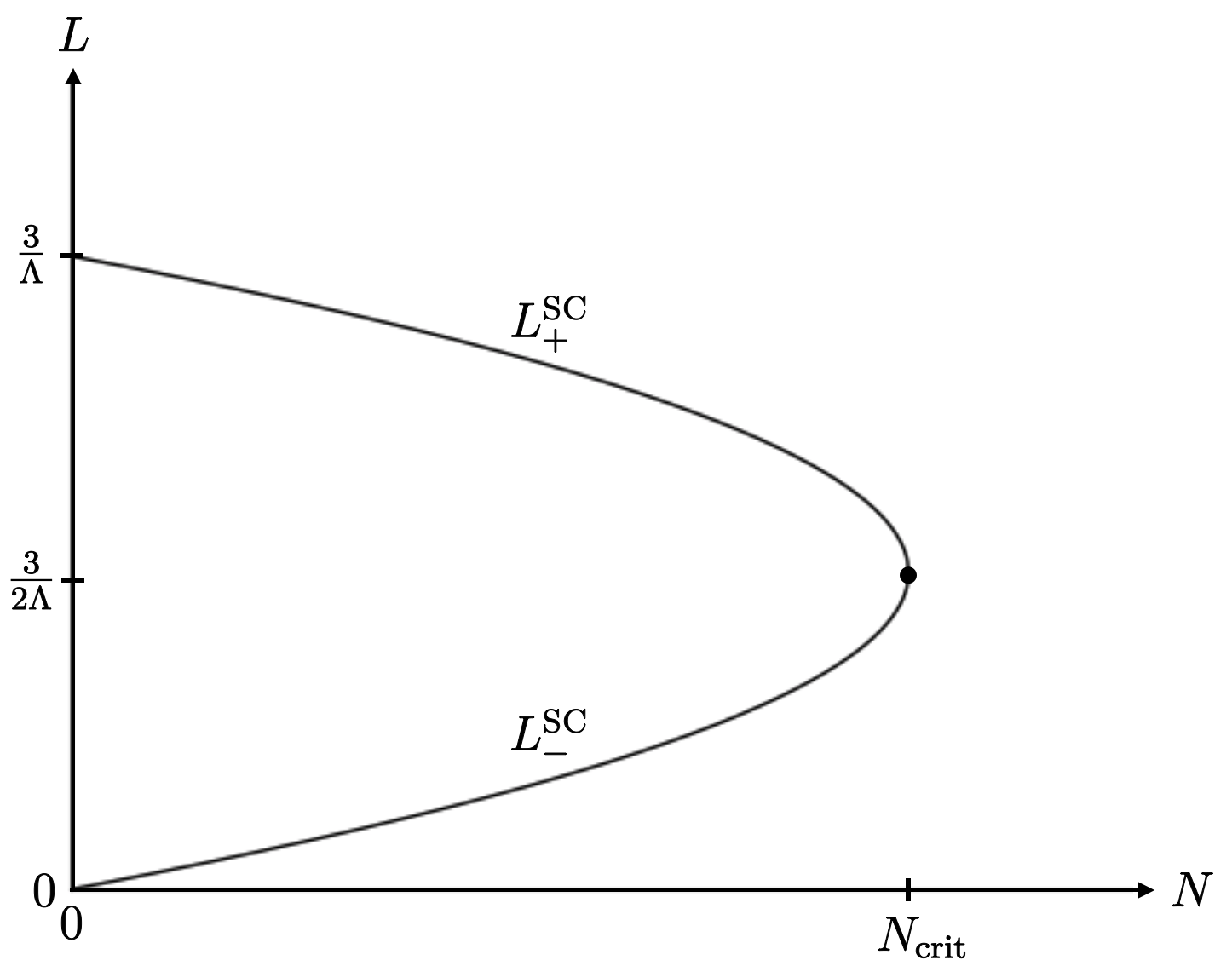}
	\caption{The two branches of self-consistent solutions $L_\pm^\text{\ac{sc}}(N)$. The upper (lower) curve corresponds to the perturbative (non-perturbative) sequence. Both sequences terminate at $N_\mrm{crit}$. No $S^4$ solutions exist beyond this point.}
	\label{fig:1}
\end{figure}

\noindent\textbf{(1) Positive bare cosmological constant.} If $\mathbf{\Lambda_\mrm{b}>0}$ this equation admits the following two branches of solutions:
\bg
\label{eq:P81}
\lef(L_\pm^\text{\ac{sc}}(N)\ri)^2=\frac{3}{2\Lambda_\mrm{b}}\lef[1\pm\sqrt{1-\frac{G\Lambda_\mrm{b}}{3\pi}\,\mathbf f(N)}\ \ri] \, .
\eg
In Figure \ref{fig:1} the radii $L_\pm^\text{\ac{sc}}(N)$ are plotted in dependence on $N$. The main properties of the two sequences are as follows.
\\ 

\noindent\textbf{(i)} In the pure gravity limit $N=0$, $\mathbf f=0$ one obtains the radii
\al{
L_+^\text{\ac{sc}}(0)&=\sqrt{3/\Lambda_\mrm{b}} \, ,\label{eq:P82}\\
L_-^\text{\ac{sc}}(0)&=0 \, ,\label{eq:P83}
}
so that the \qut{plus} branch reproduces the result from General Relativity, while the \qut{minus} branch has a singular $N=0$ limit, which corresponds to a degenerate geometry with infinite curvature.\\

\noindent\textbf{(ii)} Letting $N=1,2,3,\cdots$, the radius $L_+^\text{\ac{sc}}(N)$ decreases, while $L_-^\text{\ac{sc}}(N)$ increases, describing a meaningful geometry now.\\ 

\noindent\textbf{(iii)} There exists a critical number $N_\mrm{crit}$ at which $L_+^\text{\ac{sc}}$ and $L_-^\text{\ac{sc}}$ become equal, and beyond which there do not exist any $S^4$-type solutions to the self-consistency conditions. This critical number of degrees of freedom inhabiting the universe is reached when
$\frac{G\Lambda_\mrm{b}}{3\pi}\,\mathbf f(N_\mrm{crit})=1$.
If $G\Lambda_\mrm{b}\ll 1$ we may use the asymptotic form of $\mathbf f(N)$ and obtain
\bg
\label{eq:P85}
N_\mrm{crit}\approx\lef(\frac{36\pi}{G\Lambda_\mrm{b}}\ri)^{1/4} \, .
\eg

\noindent\textbf{(iv)} Sending $\Lambda_\mrm{b}\searrow 0$ at fixed $N$, the point where the two branches meet moves out to infinity, both in the $L$ and the $N$ direction. In this limit the \qut{minus} branch becomes
\bg
\label{eq:P88}
\lim_{\Lambda_\mrm{b}\searrow 0} L_-^\text{\ac{sc}}(N)=\lef[\frac{G}{4\pi}\,\mathbf f(N)\ri]^{1/2} \, ,
\eg
and so it reproduces exactly the solution found in the previous subsection.\\

\noindent\textbf{(v)} To summarize: For nonzero, positive $\Lambda_\mrm{b}$ there exist two $N$-sequences, a \qut{perturbative} one, $L_+^\text{\ac{sc}}(N)$, and a \qut{nonperturbative} one, $L_-^\text{\ac{sc}}(N)$.\\ \indent
Along the \textbf{perturbative sequence}, the self-consistent radius $L_+^\text{\ac{sc}}(N)$ starts out from its classical ($\hbar=0$) value, decreases with the number of degrees of freedom, and causes the curvature $R_-^\text{\ac{sc}}(N)=12/L_-^\text{\ac{sc}}(N)$ to increase correspondingly. As long as $N<N_\mrm{crit}$, this behavior is similar to the one we are familiar with from the Pauli-type calculations.\\ \indent
The \textbf{nonperturbative $L_-^\text{\ac{sc}}(N)$ sequence} instead, while having a singular initial point at $N=0$, describes a series of universes which become increasingly larger and more \qut{inhabitable} when further degrees of freedom are added.\\ \indent
Both sequences terminate at the same critical number of modes living on the approximant's spacetime. Beyond $N_\mrm{crit}$ self-consistent geometries, if they exist, are necessarily more complicated than round spheres. Qualitatively this picture is in accord with indirect evidence from the functional renormalization group indicating that vacuum fluctuations on a certain length scale generate curvature structures \ita{on that particular scale} \cite{Pagani:2019vfm,BIvac-2}.\footnote{The picture is also in line with a purely classical proposal of \qut{hiding} the cosmological constant at short scales \cite{Carlip:2018zsk}. See also \cite{Wang:2017oiy} and \cite{Hollands:2004xv} for related work in the context of \ac{qft}.}\\

\noindent\textbf{(2) Negative bare cosmological constant.} It is quite remarkable that even if $\mathbf{\Lambda_\mrm{b}<0}$ the condition \gl{P80} admits a (single) sequence of self-consistent spacetimes. As the classical Einstein equation possesses no $S^4$ solutions in this case, the sequence owes its existence entirely to the quantum effects:
\bg
\label{eq:P90}
L^\text{\ac{sc}}(N)^2=\frac{3}{2|\Lambda_\mrm{b}|}\lef[\sqrt{1+\frac{|\Lambda_\mrm{b}| G}{3\pi}\,\mathbf f(N)}-1\ri] \, .
\eg
This sequence begins at $N=1$, then the radius $L^\text{\ac{sc}}(N)$ grows monotonically with $N$, the curvature decreases, and for $N\to\infty$ the self-consistent spacetime approaches flat space finally. Clearly this behavior is a perfect surprise from both the classical gravity and the background dependent \ac{qft} point of view.

\subsection{Massive and nonminimally coupled scalars}

Up to now we assumed a vanishing bare mass, and as a consequence the integrated trace of the stress tensor was essentially the mode counting function, $\Theta_N=-\mathbf f(N)$. In the general case with $M\neq 0$ a partial fraction decomposition of \gl{P51} with \gl{P52} expresses $\Theta_N$ in the following suggestive form:
\bg
\label{eq:P95}
\spl{
-\Theta_N(L)=\mathbf f(N)+\frac{1}{6}(ML)^2\Bigg[&\,\sum_{n=1}^N(2n+3)\\
&+2(2-z)\sum_{n=1}^N\frac{n}{n(n+3)+z}\\
&+3(2-z)\sum_{n=1}^N\frac{1}{n(n+3)+z}\,\Bigg] \, .
}
\eg
Here we introduced the abbreviation $z\equiv (ML)^2+12\xi$.\\ \indent 
The leading large-$N$ behavior of the terms in the first, second and third line of \gl{P95} is, respectively, $\propto N^4$, $N^2$ and $\log(N)$; the last term, containing a convergent sum, becomes $N$ independent asymptotically. Hence a nonzero $M$ has comparatively little impact on $\Theta_N$. When $N\gg 1$, it does not modify the dominant $N^4$ and $N^3$ terms coming from $\mathbf f(N)$, and shows up at the next-to-next-to leading order only, being proportional to $(ML)^2\sum_{n=1}^N(2n+3)\propto(ML)^2N^2$. The first term in which $\xi$ makes its appearance is even more suppressed.\\ \indent
Moreover, taking the limit $ML\to\infty$ directly in \Gl{P51} at fixed $N$, we see that even within the whole span of masses between $M=0$ and $M\to\infty$ the trace $\Theta_N(L)$ changes only by a modest factor of $2$:
\bg
\label{eq:P96}
\Theta_N(L)\big|_{M\to\infty}=2\,\Theta_N(L)\big|_{M=0} \, .
\eg
Because of their weak and qualitatively unimportant role when $N\gg 1$ we shall not discuss the case of a nonzero mass and non-minimal coupling any further.

\section{The Path Integral Route}
\label{sec:path-int}

In this section we consider the coupling of the scalar field degrees of freedom to gravity from an effective action and functional integral point of view. At first sight it might seem that not much new can be learned in this way, dealing with a simple Gaussian theory after all. However, the specific needs of the present investigation bring certain aspects to light which are important now, while they often can be brushed away in typical background dependent calculations.\\ \indent 
For example, generically a scalar plus gravity system is described by an effective action $\Gamma[A;g]$ depending on arbitrary fields $A(x)$ and $g\mn(x)$. However, if one is interested in the \qut{particle physics} of $A$ on a rigid, say flat, spacetime only, there is no need to know the $A$-independent terms of $\Gamma$, i.e., the invariants constructed from $g\mn$ alone. For us those terms are crucial though.\\ \indent
So the plan of this section is, first, to pin point precisely which parts of the functional integral for $\Gamma$ do, or do not depend on the metric, and second, install a regulator that complies with the general requirements of an $N$-cutoff.\pagebreak\\ \indent
The relevant object is the induced gravity action $\Gamma_\mrm{scal}[g]$ which is generated by the vacuum fluctuations of a scalar field. In the case at hand the latter is governed by the classical action $S[A;g]\equiv\foh\int_\mM\!\dd x\sgo\,A\,\mK A$ with $\mK$ as in \gl{P2}. The functional $\Gamma_\mrm{scal}[g]$ augments the Einstein-Hilbert action of pure classical gravity
\bg
\label{eq:Q0}
S_\mrm{EH}[g]=\frac{1}{16\pi G}\int\!\dd x\sgo\,\lef(-R+2\Lambda_\mrm{b}\ri)
\eg
to the effective gravitational action
\bg
\label{eq:Q1}
\Gamma[g]=S_\mrm{EH}[g]+\Gamma_\mrm{scal}[g] \, .
\eg\indent
The functional $\Gamma[g]$ is the restriction of the full Legendre effective action $\Gamma[A;g]$ to a vanishing scalar field, $\Gamma[g]\equiv\Gamma[A;g]\big|_{A=0}$. As $A$ has no self-interactions, $\Gamma[A;g]=S[A;g]+\Gamma_\mrm{scal}[g]+S_\mrm{EH}[g]$, and therefore the effective field equation $\delta\Gamma[A;g]/\delta A(x)=0$ for the expectation value $A\equiv\langle\what A\rangle$ is given by \gl{P3}. For any metric, this equation possesses the solution $A(x)=0$, the vacuum case on which we focus throughout. As defined above, the matter action $S$ vanishes on this vacuum configuration\footnote{However, had we declared (or had Nature told us) that the correct description of the matter system necessitates the modified scalar action $S'\equiv S[A;g]+\Delta S[g]$, then the equation for $A$ remains the same, while $\Delta S[g]$ makes an additional contribution to the \ac{rhs} of \gl{Q1}.}, $S[0;g]=0$, which brings us back to \gl{Q1} for $\Gamma[0;g]\equiv\Gamma[g]$.\\ \indent
Writing down the corresponding field equation $\delta\Gamma[g]/\delta g\mn=0$, we recover essentially the semiclassical Einstein equation \gl{P30}, but with the expectation value $\langle T\mn[\what A;g]\rangle^g$ now replaced by the stress tensor $T_\Gamma\MN$:
\bg
\label{eq:Q2}
T\MN_\Gamma[g](x)\equiv-\frac{2}{\sgo}\frac{\delta}{\delta g\mn(x)}\Gamma_\mrm{scal}[g] \, .
\eg\indent
One of the questions we are going to address is the precise relationship between these two candidates for a semiclassical stress tensor, and what the implications are for the existence of $\{\mathsf{App}(N)\}$-sequences.

\subsection{Functional integral and measure}

\enlargethispage{\baselineskip}
We start out from the general case of arbitrary $d$-dimensional, compact Euclidean spacetimes $(\mM,g)$ without boundary. The functional integral representation of the induced gravity action due to a scalar field on such spacetimes reads then \cite{Toms:1987,Fujikawa:2004cx,Unz:1985wq}
\bg
\label{eq:Q3}
\e^{-\Gamma[g]}=\intD(A;g)\,\e^{-S[A;g]} \, .
\eg
The measure $\mD(A;g)$ brings in an extra metric dependence. It has the form, for any dimensionality $d$,
\bg
\label{eq:Q4}
\mD(A;g)=\prod_x\lef[\det(g\mn(x))\ri]^{1/4}\mu\,\D A(x) \, .
\eg
The mass parameter $\mu$ is included here to make $\mD(A;g)$ dimensionless.\pagebreak\\ \indent
For the purposes of this section we assume that the functional integral \gl{Q3} has been regularized by restricting it to a finite number of spacetime points, and attaching integration variables $A(x)$ only to the sites $x$ of, say, a lattice or a triangulation. The details of this preliminary regularization are not important though and we do not make them explicit.\\ 
\noindent\textbf{(1)} The integral \gl{Q3} with \gl{Q4} is the Euclidean analog of the quantum mechanical path integral which is strictly equivalent to applying the rules of canonical quantization. Its derivation starts out from the operatorial formalism, then uses discretization techniques to construct a Hamiltonian functional integral involving a generalized Liouville measure for paths on phase space, and finally performs the Gaussian integration over the field momenta to arrive at its Lagrangian version \cite{Toms:1987,Unz:1985wq}. The same result is obtained by arguments based upon \ac{brst} invariance \cite{Fujikawa:2004cx}.\\ \indent
It must be emphasized that the metric dependence of the above functional measure is by no means anything \qut{exotic}. In fact, it is the path integral with just \ita{this} measure which underlies the well known trace-log formula abundantly used in one-loop computations.\\ 
\noindent\textbf{(2)} Let us now perform the integration over $A(x)$ in \gl{Q3}, thereby paying careful attention to the possibility of hidden metric dependencies. Given a fixed metric $g\mn$, the usual procedure consists in first diagonalizing the associated kinetic operator $\mK\equiv \mK[g]$, then expanding the field in terms of its eigenfunctions, $A(x)=\sum_{n,m}\alpha_{n,m}u_{n,m}(x)$, and finally changing from the integration variables $A(x)$, for all $x$, to the set of all $\alpha_{n,m}$. One might be suspicious that perhaps some hidden metric dependence creeps in during this procedure, in particular since the $u_{n,m}$s satisfy orthogonality and closure relations that do depend on the metric.\\ \indent
To show that this is not the case actually we first rewrite the functional integral in terms of the new field
\bg
\label{eq:Q10}
B(x)\equiv g^{1/4}(x)A(x) \, .
\eg
It transforms as a scalar density, and has the welcome property that it renders the transformed measure metric independent:
\bg
\label{eq:Q11}
\spl{
\int\!\prod_x g^{1/4}(x)\,\D A(x)&\,\exp\lef\{-\foh\int\!\dd x\, g^{1/2}\, A\mK A\ri\}\\
&=\int\!\prod_x\D B(x)\,\exp\lef\{-\foh\int\!\dd x\, B\widetilde{\mK} B\ri\} \, .
}
\eg
In the integral over $B(x)$, the entire metric dependence resides in the new kinetic operator
\bg
\label{eq:Q12}
\widetilde\mK\equiv g^{1/4}(x)\,\mK\,g^{-1/4}(x) \, .
\eg\indent
Furthermore, we introduce a new set of basis functions $\{v_{n,m}\}$ defined as densitized versions of the $u_{n,m}$s:
\bg
\label{eq:Q15}
v_{n,m}(x)\equiv g^{1/4}(x)\,u_{n,m}(x) \, .
\eg
According to Eqs. \gl{P10-1} and \gl{P10-2} they enjoy the properties
\al{
\int\!\dd x\,v_{n,m}^*(x)\, v_{\bar n,\bar m}(x)&=\delta_{n\bar n}\,\delta_{m\bar m}\label{eq:Q16}\\
\sum_{n,m}v_{n,m}(x)\,v_{n,m}^*(y)&=\delta(x-y) \, .\label{eq:Q17}
}
These relations do not involve the metric any longer. Most importantly, the $u$s being eigenfunctions of $\mK(g)$ implies that the $v$s are eigenfunctions of $\widetilde\mK$, with the same eigenvalues:
\bg
\label{eq:Q18}
\widetilde\mK[g]\,v_{n,m}(x)=\mF_n\, v_{n,m}(x) \, .
\eg
Being related by a similarity transformation, $\mK$ and $\widetilde\mK$ have identical spectra.\\ \indent
Taking advantage of the $v$-basis we can now perform the $B$-integral of \gl{Q11} in a completely clearcut manner. After expanding the field as
\bg
\label{eq:Q19}
B(x)=\sum_{n,m}b_{n,m}\,v_{n,m}(x)
\eg
we change integration variables from $B(x)$ to $b_{n,m}$. The relations \gl{Q16} and \gl{Q17} imply that the corresponding Jacobian matrix ${J_x}_{n,m}\equiv\p B(x)/\p b_{n,m}=v_{n,m}(x)$ is orthogonal formally, and has unit determinant therefore.\footnote{The conditions of orthogonality, $J^\dagger J=J J^\dagger=\mathds{1}$, are easily seen to be nothing but the orthonormality and completeness relations of the $v$s in disguise. Suppressing the degeneracy indices for clarity, one has indeed, at the formal level, $(J^\dagger J)_{n\bar n}=\int\!\dd x\,(J^\dagger)_{nx}J_{x\bar n}=\int\!\dd x J^*_{xn}J_{x\bar n}=\int\!\dd x\,v_n^*(x)v_{\bar n}(x)=\delta_{n\bar n}\equiv\mathds{1}_{n\bar n}$, as well as $(J J^\dagger)_{xy}=\sum_n J_{xn}(J^\dagger)_{ny}=\sum_n J_{xn}J^*_{yn}=\sum_n v_n(x)v_n^*(y)=\delta(x-y)\equiv\mathds{1}_{xy}$.}\\ \indent
Thus the integral \gl{Q11} boils down to
\bg
\label{eq:Q22}
\prod_{n,m}\int\!\D b_{n,m}\,\exp\lef\{-\foh\sum_{n,m}\mF_n\, b_{n,m}^2\ri\}
\eg
and hence, up to an irrelevant numerical constant,
\bg
\label{eq:Q23}
\exp\lef\{-\Gamma_\mrm{scal}[g]\ri\}=\lef(\prod_{n,m}\frac{\mF_n}{\mu^2}\ri)^{-1/2} \, .
\eg
This brings us to the (expected, of course) final result:
\bg
\label{eq:Q24}
\spl{
\Gamma_\mrm{scal}[g]&=\foh\sum_{n,m}\log(\mF_n/\mu^2)\\
&=\foh\sum_n D_n\log(\mF_n/\mu^2) \, .
}
\eg

\noindent\textbf{(3)} The careful derivation we just went through highlights several points which are particularly relevant here.\\

\noindent\textbf{(i)} The representation \gl{Q24} of the induced gravity action makes it manifest that $\Gamma_\mrm{scal}$ depends on the metric exclusively via the spectral data of the operators $\widetilde\mK$, or what amounts to the same, $\mK$. Importantly, this property emerges only thanks to the presence of the explicit $g^{1/4}$-factors in the measure $\mD(A;g)$.\\

\noindent\textbf{(ii)} It is tempting to write \gl{Q24} in the style of an operator trace,
\bg
\label{eq:Q25}
\Gamma_\mrm{scal}[g]=\foh\Tr\log(\mK/\mu^2)=\foh\Tr\log(\widetilde\mK/\mu^2) \, .
\eg
We shall refrain from this formal notation however because it tends to obscure things again:\\ \indent 
When the lattice cutoff which we tacitly invoked up to here is lifted, the trace must be regularized in some other way, $\Tr\to\Tr_\mrm{reg}$, and depending on how this is done, further, unintended metric dependencies may creep in. Moreover, with a generic regularization, \qut{$\Tr_\mrm{reg}$} might fail to satisfy all defining properties of a trace. If so, given the relation \gl{Q12} between $\mK$ and $\widetilde\mK$, one may have difficulties in establishing the second equality of \gl{Q25}, or it is violated even.\\ \indent 
Similar remarks apply to the naively equivalent $\Gamma_\mrm{scale}[g]=\foh\log\Det(\mK/\mu^2)$. Here the regularization can destroy the general properties of a determinant by multiplicative anomalies. While working at finite $N$ such difficulties will not concern us now.

\subsection{Induced gravity action with $N$-cutoff}

Within the framework advocated in this paper, one sidesteps the problems raised at the end of the previous subsection by thinking of $\Gamma_\mrm{scal}[g]$ as a quantity whose sole input information is the spectrum of $\mK$, being explicitly given by \Gl{Q24}. To be in line with our earlier discussion we now lift the lattice-type cutoff that was implicitly behind our derivation, and we replace it by an $N$-cutoff.\\

\noindent\textbf{(1) The $N$-cutoff.} At this point we return to the maximally symmetric example $\mM=S^d(L)$ and compute the correspondingly restricted functional $\Gamma_\mrm{scal}[L^2\gamma\mn]\equiv\Gamma_\mrm{scal}(L,N)$ by equipping \gl{Q24} with the same type of $N$-cutoff as in the previous sections. Truncating the summation over the $\Box_g$-eigenvalues at $n=N$, \Gl{Q24} becomes
\bg
\label{eq:Q29}
\Gamma_\mrm{scal}(L,N)=\foh\sum_{n=1}^N D_n\log\Big([\mE_n(L)+\xi R(L)+M^2]/\mu^2\Big) \, .
\eg
Specifically when $d=4$, 
\bg
\label{eq:Q30}
\Gamma_\mrm{scal}(L,N)=\foh\sum_{n=1}^N D_n\log\lef(\frac{n(n+3)+12\xi+(ML)^2}{(\mu L)^2}\ri) \, .
\eg
Recalling that $\sum_{n=1}^N D_n\equiv\mathbf f(N)$, it is convenient to rewrite \gl{Q30} in the form
\bg
\label{eq:Q31}
\Gamma_\mrm{scal}(L,N)=-\log(\mu L)\,\mathbf f(N)+\Delta\Gamma(ML,N)
\eg
where
\bg
\label{eq:Q32}
\Delta\Gamma(ML,N)\equiv\foh\sum_{n=1}^N D_n\log\lef[n(n+3)+12\xi+(ML)^2\ri] \, .
\eg\indent
Simple as it looks, the effective action \gl{Q31} is quite remarkable and surprising. Let us specialize for massless scalars for a moment, 
\bg
\label{eq:Q35}
\Gamma_\mrm{scal}(L,N)=-\mathbf f(N)\,\log(\mu L)+\Delta\Gamma(0,N)\quad ,\quad(M=0) \, .
\eg
After having set $M=0$ in \gl{Q32}, the contribution $\Delta\Gamma(0,N)$ is seen to be perfectly \ita{independent of $L$}. Thus we conclude that the \ita{exact} $L$-dependence of the induced action is of the form
\bg
\label{eq:Q36}
c_1(N)+c_2(N)\,\log(L)
\eg
with $N$-dependent constants $c_{1,2}$. Asymptotically, $c_{1,2}\propto N^4$.\\ \indent
This $L$-dependence is one of our main results. In particular we stress that, contrary to general expectations, \ita{no terms proportional to $L^4$ or $L^2$ are induced.}\\

\noindent\textbf{(2) $\mathscr P$-cutoffs.} Many of the traditional calculations with a dimensionful \ac{uv} cutoff at $\mathscr P$ would instead of \gl{Q36} produce a structure like, omitting prefactors,
\bg
\label{eq:Q37}
\mathscr{P}^4 L^4+\mathscr{P}^2 L^2+\cdots \, .
\eg
It descends from the first terms of the general action \cite{Birrell:1982ix, Parker:2009uva}
\bg
\label{eq:Q38}
\Gamma_\mrm{scal}[g]=\mathscr{P}^4\int\!\D^4 x\sgo+\mathscr{P}^2\int\!\D^4 x\sgo\, R+\cdots
\eg
obtained from a derivative expansion, or by employing the asymptotic heat kernel series for early proper times $s$ and identifying $\mathscr P$ with $1/\sqrt{s}$ there. Flat space approaches based upon plane waves and a standard momentum cutoff $p\m^2\leq\mathscr P^2$ also led to \gl{Q38}. While the first few terms of the series \gl{Q38} diverge for $\mathscr P\to\infty$, they involve invariants already present in the Einstein-Hilbert action (possibly generalized by higher derivative terms). Hence the divergences can be absorbed by redefinitions of parameters like $G$ and $\Lambda_\mrm{b}$.\\

\noindent\textbf{(3) No quartic (quadratic) renormalization of the cosmological (Newton) constant.} The potential significance of the $N$-cutoffs advocated here is understood best by comparing \gl{Q37} to \gl{Q36}. When the dimensionful cutoff $\mathscr P$ is employed, the general structure of \gl{Q37}, and more generally of $\Gamma_\mrm{scal}[g]$ in \gl{Q38}, is fixed to a very large extent by simple dimensional analysis.\\ \indent 
When no other dimensionful parameter is available, invariants with mass dimension $-k$ cannot but get multiplied by a prefactor proportional to $\mathscr{P}^k$, \ita{whatever are the details of the concrete calculation}. In particular it is unavoidable that the invariants $\int\!\D^4 x\sgo$ and $\int\!\D^4 x\sgo R$ arise multiplied by $\mathscr P^4$ and $\mathscr P^2$, respectively.\\ \indent 
While the interpretation of these $\mathscr P$-dependencies is somewhat different in fundamental and effective theories, they always seem to indicate the presence of strong quantum effects that try to change the values of the cosmological and Newton's constant. However, those effects can very well be a pure artifact of the formalism employed, namely a dimensionful cutoff plus an asymptotic expansion. Being virtually unavoidable, there is no guarantee that the $\mathscr P^k$-dependencies reflect any real physical effect that occurs as a result of specific dynamical assumptions about the system.\\ \indent
$N$-type cutoffs, on the other hand, being dimensionless, are free from this kind of prejudice about the cutoff dependence of the invariants. Therefore it can be expected that the $N$-dependences they give rise to are more likely to contain genuine physics information than the standard $\mathscr P$-dependences.\\ \indent
The absence of a $\mathscr{P}^4$ term in the exact result \gl{Q36} indicates already that the cosmological constant issue will present itself differently here; we shall come back to it in more detail in Subsection \ref{subsec:complete}.

\subsection{Semiclassical stress tensors: a second candidate}

Now we turn to the effective Einstein equation implied by the stationarity of $S_\mrm{EH}[g]+\Gamma_\mrm{scal}[g]$. Its traced and integrated form with $S^d$ metrics inserted has the same structure as in the previous section,
\bg
\label{eq:Q50}
-\foh(d-2)\,R(L)+d\,\Lambda_\mrm{b}=8\pi G\,\frac{\Theta_N^\Gamma(L)}{\mrm{Vol}[S^d(L)]}
\eg
but now the backreaction of the quantum system is controlled by the quantity
\bg
\label{eq:Q51}
\Theta_N^\Gamma(L)\equiv\int\!\dd x\sgo\ {T_\Gamma}\m\M[g](x)\equiv\mT\Gamma_\mrm{scal}[g] \, .
\eg
It involves the stress tensor \gl{Q2} and the metric $g\mn=L^2\gamma\mn$.\\ \indent 
In \Gl{Q51} we introduced the derivative operator
\bg
\label{eq:Q52}
\mT\equiv-2\int\!\dd x\,g\mn(x)\frac{\delta}{\delta g\mn(x)}
\eg
which we must apply to $\Gamma_\mrm{scal}$ from \Gl{Q29} now. To do so we use the following handy and generally valid rule which is easily derived by a functional Taylor expansion.\\ \indent 
Let $F\equiv F[g\mn]$ be an arbitrary functional and ${T_F}\MN\equiv-\frac{2}{\sgo}\frac{\delta F}{\delta g\mn}$ the associated stress tensor. Then $\mT F[g]\equiv\int\!\dd x\,{T_F}\m\M[g](x)$ is given by
\bg
\label{eq:Q55}
\mT F[g] =\frac{\D}{\D\alpha}F\lef[\e^{-2\alpha}g\mn\ri]\bigg|_{\alpha=0}
\eg
where $\alpha$ has the interpretation of a position independent Weyl parameter. Furthermore, if $F(L)\equiv F[L^2\gamma\mn]$ denotes the restriction of $F$ to metrics on $S^d$, the operator $\mT$ acts on such functions of the radius according to 
\bg
\label{eq:Q56}
\mT F(L)=-L\frac{\D}{\D L}F(L) \, .
\eg\indent
Upon applying this rule to \gl{Q29} we obtain, for any dimensionality $d$,
\bg
\label{eq:Q57}
\Theta_N^\Gamma(L)=\sum_{n=1}^N D_n\lef[1-\frac{M^2}{\mE_n(L)+\xi R(L)+M^2}\ri] \, .
\eg
The following points should be noted here.\\

\noindent\textbf{(1)} The arbitrary mass scale $\mu$ has dropped out from \gl{Q57} and the effective Einstein equation.

\noindent\textbf{(2)} For every fixed value of $N$ one has the following, both $d$ and $L$ independent limiting values for small and large masses, respectively:
\al{
\Theta_N^\Gamma(L)\big|_{M=0}&=\mathbf f(N)\label{eq:Q58}\\
\Theta_N^\Gamma(L)\big|_{M\to\infty}&=0\label{eq:Q59}
}
Obviously very heavy scalars \qut{decouple} and do not modify Einstein's equation. This kind of decoupling did not take place with the first stress tensor candidate, see \Gl{P49}.\\

\noindent\textbf{(3)} Indeed, $\Theta_N^\Gamma$ should be contrasted with its cousin
\bg
\label{eq:Q60}
\Theta_N(L)\equiv\lef\langle\mT S[\what A;g]\ri\rangle_N^g
\eg
which we computed in the previous section by straightforwardly evaluating expectation values. Comparing \gl{Q57} to \gl{P43} reveals that the integrated traces differ in all dimensions by a $M$- and $\xi$-independent term:
\bg
\label{eq:Q61}
\mT\Gamma_\mrm{scal}[g]-\lef\langle\mT S[\what A;g]\ri\rangle_N^g=\frac{d}{2}\ \mathbf f(N) \, .
\eg
As we are going to demonstrate in the next subsection, \ita{this difference is due to the metric dependence of the measure $\mD(A;g)$}.\\ \indent 
The special case $M=0$ makes it particularly clear that the two candidates for a quantum mechanical stress tensor entail semiclassical Einstein equations with quite different properties possibly.  In fact,
$\Theta_N^\Gamma\big|_{M=0}=\mathbf f(N)$
is always positive, while
$\Theta_N\big|_{M=0}=-\foh(d-2)\,\mathbf f(N)$
is negative for all $d>2$.

\subsection{The contribution from the functional measure}

In order to understand the difference between $\Theta_N$ and $\Theta_N^\Gamma$ from first principles we return to the discretization-regularized functional integral \gl{Q3} and its generalization for arbitrary expectation values,
\bg
\label{eq:Q70}
\lef\langle\mO(\what A)\ri\rangle^g\equiv\e^{+\Gamma_\mrm{scal}[g]}\intD(A;g)\,\mO(A)\,\e^{-S[A;g]} \, .
\eg
They are normalized such that $\langle\mathds{1}\rangle^g=1$ for all $g\mn$. (Indeed, the background metric is left arbitrary in this subsection.) Apart from the type of regularization, the expectation values evaluated in Section \ref{sec:first-type} are of this sort; in particular the 2-point function $\langle A(x)A(y)\rangle^g\equiv G(x,y)$ and the stress tensor trace $\langle\mT S[\what A;g]\rangle^g\equiv\Theta_N$ are examples of \gl{Q70}.\\

\noindent\textbf{(1) A general identity.} Let us now apply the derivative operator $\mT$, the generator of global Weyl transformations of the metric, to both sides of \Gl{Q3}. We obtain
\bg
\label{eq:Q75}
\e^{-\Gamma_\mrm{scal}[g]}\,\mT\Gamma[g]=\intD(A;g)\,\mT S[A;g]\,\e^{-S[A;g]}-\int\!\e^{-S[A;g]}\mT\mD(A;g) \, .
\eg
In the last term, the $g\mn$-derivative acts upon the metric dependence of the measure,
\bg
\label{eq:Q76}
\mT\mD(A;g)=\lef(\prod_{x'}\D A(x')\ri)\cdot\mT\prod_x g^{1/4}(x) \, .
\eg
On the discrete spacetime the product over $x$ is well defined, and applying the rule \gl{Q55} to it yields
\bg
\label{eq:Q77}
\spl{
\mT\prod_x g^{1/4}(x)&=\lim_{\alpha\to 0}\,\frac{\D}{\D\alpha}\,\prod_x\det{}^{1/4}\lef(\e^{-2\alpha}g\mn\ri)\\
&=\lim_{\alpha\to 0}\,\frac{\D}{\D\alpha}\,\prod_x\lef\{\e^{-d\alpha/2}\det{}^{1/4}(g\mn(x))\ri\}\\
&=\lim_{\alpha\to 0}\,\frac{\D}{\D\alpha}\,\exp\lef\{-\sum_x\frac{d\alpha}{2}\ri\}\prod_{x'}g^{1/4}(x')\\
&=\lef(-\frac{d}{2}\sum_x 1\ri)\,\prod_{x'}g^{1/4}(x') \, .
}
\eg
Thus the response of the measure to the $\mT$-transformation consists essentially in a multiplication by the number of spacetime points the regularized functional integral is based upon:
\bg
\label{eq:Q78}
\mT\mD(A;g)=\lef(-\frac{d}{2}\sum_x 1\ri)\cdot\mD(A;g) \, .
\eg
Hence using \gl{Q78} in \gl{Q75} we obtain the Ward identity we wanted to derive:
\bg
\label{eq:Q79}
\mT\Gamma_\mrm{scal}[g]-\lef\langle\mT S[\what A;g]\ri\rangle^g=\frac{d}{2}\lef(\sum_x 1\ri) \, .
\eg
This relationship has the same structure as the equation \gl{Q61} which we had discovered before by an explicit calculation.\pagebreak\\ \indent
Moreover, the two relations are \ita{strictly identical} even, since as long as both the preliminary discretization-based cutoff and the continuum $N$-cutoff are in place, it holds true that
\bg
\label{eq:Q80}
\lef(\sum_x 1\ri)\,=\ \sum_{n=1}^N D_n\equiv \mathbf f(N) \, .
\eg
The argument here is the familiar one \cite{feynman2010quantum}: The point of contact between the discretization-based cutoff and the $N$-cutoff is the expansion \gl{Q19}; it connects the integration variables employed by the former, namely $\{B(x_j)\ |\ j=1,2,\cdots,\sum_x 1\}$ where the $x_j$s are the coordinates of the lattice points, to those of the latter, the expansion coefficients $\{b_{n,m}\ |\ n=1,\cdots,N\ ;\ m=1,\cdots,D_n\}$. Since the linear relations $B(x_j)=\sum_{n,m}b_{n,m}v_{n,m}(x_j)$, $j=1,2,\cdots$, establish an invertible map between the two sets of variables it is clear that they must be equal in number, and this is what proves \gl{Q80}.\\ 

\noindent\textbf{(2) Inequivalent stress tensor candidates.} In summary, the conclusion about the self-gravitating systems $\mathsf{App}(N)$ constructed in Section \ref{sec:first-type} and in the present section, respectively, is that they may well describe different physics since they are based upon inequivalent stress tensors in the effective Einstein equation. The $N\to\infty$ limits of such approximants might differ correspondingly.\\ \indent
In Section \ref{sec:first-type} we followed a kind of \qut{bottom up} approach. It used the classical stress tensor $T\MN\propto\delta S/\delta g\mn$ as an inspiration for postulating a quantum mechanical observable ${\what T\M}\m$, and computed its expectation value by appropriately differentiating the 2-point function. Technically this method appears closer to first quantization than to \ac{qft}.\\ \indent
In the present section, on the other hand, the approach is \qut{top down}, as its inspiration for the stress tensor comes from the induced gravity action of a \ac{qft} already. While only formal, upon equipping it with an $N$-cutoff this \ac{qft} action leads to well defined self-gravitating approximants. (The corresponding sequences will be discussed in Subsection \ref{subsec:complete} below.)\\ \indent
It goes without saying that ultimately only experiment or additional theoretical criteria can decide about the correct stress tensor. Like in the renormalization of operator products that are plagued by short distance singularities, further input is needed.\\

\noindent\textbf{(3) Heat kernel regularization.} In Eqs. \gl{Q79} and \gl{Q80} we encountered the formal sum \qut{$\sum_x 1$} which, by our regularization, is assigned the value $\sum_{n=1}^N D_n\equiv\mathbf f(N)$. Concerning this crucial sum, it is instructive again to compare the $N$-cutoff to other types of cutoffs used in the literature.\\ \indent 
We focus on the heat kernel cutoff here. It can arise in a variety of ways, Fujikawa's approach to anomalies\footnote{While at first sight the above derivation of the measure contribution is reminiscent of Fujikawa's anomaly calculation \cite{Fujikawa:2004cx}, it must be stressed that, when $M=0$, in our computation \ita{all} contributions to the trace of the stress tensor are proportional to $\sum_x 1=\sum_n D_n$, hence anomaly- and non-anomalous terms appear mixed. For their disentangled form, and a detailed discussion of the Weyl-Ward identities we refer to \cite{N-2}.} being a well known example \cite{Fujikawa:2004cx,Bastianelli:2006rx}. There are characteristic differences between those two regularization schemes which further highlight the inequivalence of $N$- and $\mathscr P$-type cutoffs.\pagebreak\\ \indent 
The heat kernel method interprets the sum $(\sum_x 1)$ as the a priori ill defined trace of the unit operator $\mathds{1}=(\delta_{xy})$, and regularizes it, strictly in the continuum, by inserting a damping factor that suppresses all contributions from modes with $\mK$-eigenvalues $\mF\gtrsim\mathscr{P}^2$; as suggested by its name, the parameter $\mathscr P$ has the dimension of a mass again:
\bg
\label{eq:Q85}
\lef(\sum_x 1\ri)_\mrm{reg}=\ \Tr[\mathds{1}]_\mrm{reg}\ \equiv\ \Tr\!\lef[\e^{-\mK/\mathscr{P}^2}\ri] \, .
\eg\indent
What used to be a pure number, $(\sum_x 1)$, has become a functional of the metric by this regularization. Here, too, the process of removing the regulator is understood to mean letting $\mathscr P\to\infty$, rather than $\mathbf f\to\infty$. For dimensional reasons the asymptotics of \gl{Q85} has the structure, in $d=4$,
\bg
\label{eq:Q86}
\lef(\sum_x 1\ri)_\mrm{reg}=\int\!\D^4 x\sgo\,\bigg\{c_4\mathscr{P}^4+c_2 R\mathscr{P}^2+(\text{curvature})^2\text{-terms}+O(\mathscr{P}^{-2})\bigg\} \, .
\eg
Explicit calculations find for the $(\text{curvature})^2$-terms the celebrated anomaly structure 
$a E_4-c C\mnab C\MNAB$
where $E_4$ is the Euler form density and $C\mnab$ the Weyl tensor \cite{Mottola:2010gp,Komargodski:2011vj}. The scheme independent coefficients $a$ and $c$ are known for many types of fields \cite{Mottola:2010gp}.\\ \indent On spheres, \Gl{Q86} yields an $L$-dependence of the form
\bg
\label{eq:Q88}
\lef(\sum_x 1\ri)_\mrm{reg}=L^4 \mathscr{P}^4+L^2\mathscr{P}^2+\mrm{const}+O(L^{-2}) \, .
\eg
Being a typical heat-kernel based result, \gl{Q88} must be contrasted with its perfectly $L$-independent counterpart obtained with the $N$-cutoff, \Gl{Q80}.

\subsection{A complete $N$-sequence with $\Lambda_\mrm{b}>0$}
\label{subsec:complete}

In this subsection we discuss the sequences $\{\mathsf{App}(N)\}$ which follow from the $4D$ effective Einstein equation
\bg
\label{eq:Q100}
-\frac{12}{L^2}+4\Lambda_\mrm{b}=\frac{3G}{\pi}\,\frac{\Theta_N^\Gamma}{L^4}
\eg
when the second candidate for the semiclassical stress tensor is employed; $\Theta_N^\Gamma(L)$ is given by \Gl{Q57} then. We restrict the discussion to the case $M=0$, implying that now $\Theta_N^\Gamma=\mathbf f(N)$, as opposed to $\Theta_N=-\mathbf f(N)$ for the first candidate in $d=4$. As a consequence, there are no solutions to \gl{Q100} with a vanishing or a negative bare cosmological constant.\\ \indent 
On the other hand, for every given positive $\Lambda_\mrm{b}\equiv 3/L_\mrm{b}^2$ there exists a complete sequence of self-consistent radii:
\bg
\label{eq:Q105}
\spl{
L^\text{\ac{sc}}(N)^2&=\frac{3}{2\Lambda_\mrm{b}}\lef[1+\sqrt{1+\frac{G\Lambda_\mrm{b}}{3\pi}\mathbf f(N)}\,\ri]\\
&\equiv\foh L_\mrm{b}^2\lef[1+\sqrt{1+\frac{G}{\pi L_\mrm{b}^2}\mathbf f(N)}\,\ri] \, .
}
\eg
The sequence of self-gravitating approximants that live on the respective spacetimes $S^4\lef(L^\text{\ac{sc}}(N)\ri)$ has a number of remarkable properties:\\ 

\noindent\textbf{(1)} The systems $\mathsf{App}(N)$ exist for all $N=0,1,2,\cdots$; the underlying spacetime $S^4\lef(L^\text{\ac{sc}}(N)\ri)$ is always nondegenerate. In particular this sequence does possess a classical initial point, $L^\text{\ac{sc}}(0)=L_\mrm{b}$, contrary to the example in Section \ref{sec:seq-self}.\\

\noindent\textbf{(2)} The self-consistent radii \gl{Q105} are a monotonically \ita{increasing} function of $N$. The universe of the approximants \ita{expands} when further quantum mechanical degrees of freedom are added.\\ \indent
This solution, too, disproves the prejudice underlying the \qut{cosmological constant problem} which maintains that vacuum fluctuations increase the effective cosmological constant, causing the universe to \ita{shrink}.\\ \indent
The equation \gl{Q105} becomes most transparent when $\mathbf f\gg 1$,
\bg
\label{eq:Q106}
L^\text{\ac{sc}}(N)^2\approx L_\mrm{b}\sqrt{\frac{G}{4\pi}\mathbf f(N)} \, .
\eg
If we also approximate $\mathbf f(N)\approx\frac{1}{12}N^4$ the formula is very simple and instructive:
\bg
\label{eq:Q107}
L^\text{\ac{sc}}(N)\approx\lef(\frac{G\hbar}{48\pi}\ri)^{1/4}L_\mrm{b}^{1/2}\, N \, .
\eg
In writing down \gl{Q107} we reinstated Planck's constant for a moment. We observe that the relation between $L^\text{\ac{sc}}$ and $N$ becomes \ita{linear} when $N\gg 1$, and that it depends on both $G$ and $\hbar$ \ita{in a non-analytic way}. This is a clear indication of its non-perturbative origin. As the $4D$ Planck length is given by $\ell_\mrm{Pl}=(\hbar G/c^3)^{1/2}$, i.e., $\ell_\mrm{Pl}=(\hbar G)^{1/2}$ in our units, we can absorb the $\hbar$-dependence of \gl{Q107} in a dependence on $\ell_\mrm{Pl}$, yielding
\bg
\label{eq:Q108}
L^\text{\ac{sc}}(N)\approx\frac{N}{(48\pi)^{1/4}}\sqrt{\ell_\mrm{Pl}\,L_\mrm{b}} \, ,
\eg
which is non-analytic in the Planck length.\\

\noindent\textbf{(3)} If we let $N\to\infty$, the \qut{Hubble} radius $L^\text{\ac{sc}}(N)$ grows unboundedly, and spacetime becomes flat \ita{for every fixed value of the bare cosmological constant $\Lambda_\mrm{b}\equiv 3/L_\mrm{b}^2$}:
\bg
\label{eq:Q110}
S^4\lef(L^\text{\ac{sc}}(N)\ri)\ \xrightarrow{N\to\infty}\ \mathcal{R}^4 \, .
\eg\indent
We summarize this behavior as follows: The quantum field theory describing a free scalar field interacting with classical gravity is defined as the limit of a sequence of approximants $\mathsf{App}(N)$ which symbolically can be written as
\bg
\label{eq:Q111}
\Big(\mathbf f(N)\ \text{scalar modes}\Big)\otimes\Big(\text{self-consistent spacetime}\Big)_N \, .
\eg
We advocate the point of view that the correct limit for removing the regulator is $N\to\infty$, rather than $\mathscr P\to\infty$ where $\mathscr P$ is any dimensionful cutoff scale. On the basis of the self-consistency condition considered (maximally symmetry restricted, second stress tensor) there exists only one such sequence, namely \gl{Q105}, and this sequence $\{\mathsf{App}(N)\}_{N\in\mathds{N}}$ converges in a well defined way to 
\bg
\label{eq:Q112}
\Big(\text{free scalar field, fully quantized}\Big)\otimes\mathcal{R}^4 \, .
\eg\indent
Thus, if one quantizes the matter field by this approach, and starts out from a classical spacetime $S^4(L_\mrm{b})$ with $L_\mrm{b}$ fixed, but arbitrary, the outcome will \ita{always} be that \ita{the fully quantized scalar field lives on a flat spacetime.} Or stated differently, flat space emerges without any finetuning.\\

\noindent\textbf{(4)} To gain further insight into the Background Independent treatment and its capability to produce a result diametrically opposite to the standard one, let us look at the dimensionful cutoff $\mathscr P$ corresponding to a given value of $N$:
\bg
\label{eq:Q120}
\mathscr{P}(N)^2\equiv\mE_N(L^\text{\ac{sc}}(N))=\frac{N(N+3)}{L^\text{\ac{sc}}(N)^2} \, .
\eg
For the solution \gl{Q105} we obtain explicitly 
\bg
\label{eq:Q121}
\mathscr{P}(N)^2=\frac{2N(N+3)}{L_\mrm{b}^2}\lef[1+\sqrt{1+\frac{12\mathbf f(N)}{N_T^4}}\,\ri]^{-1} \, ,
\eg
with the abbreviation
\bg
\label{eq:Q122}
N_T\equiv(12\pi)^{1/4}\lef(\frac{L_\mrm{b}}{\ell_\mrm{Pl}}\ri)^{1/2} \, .
\eg
Therefore, when $N\gg 1$, we have in units of $m_\mrm{Pl}\equiv\ell_\mrm{Pl}^{-1}$,
\bg
\label{eq:Q123}
\mathscr{P}(N)^2=(24\pi)\,m_\mrm{Pl}^2\,\frac{N^2}{N_T^4}\lef[1+\sqrt{1+\lef(\frac{N}{N_T}\ri)^4}\,\ri]^{-1} \, .
\eg
\begin{figure}[ht]
	\centering
  \includegraphics[width=0.7\textwidth]{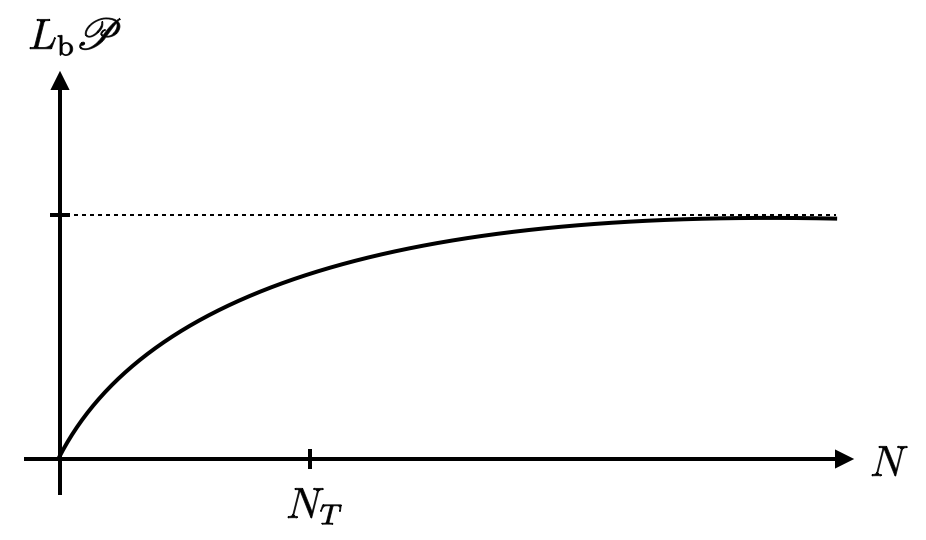}
	\caption{The dimensionful cutoff scale $\mathscr P$ in dependence on $N$ according to \Gl{Q121}. The corresponding sequence of approximants assumes a positive bare cosmological constant.}
	\label{fig:2}
\end{figure}
Figure \ref{fig:2} shows a graph of the function $\mathscr{P}(N)$ for $N_T\gg 1$. As expected, the dimensionful \ac{uv} cutoff scale is zero at the classical point, $\mathscr P(N=0)=0$, then increases as we let $N=1,2,3,\cdots$, but as soon as $N$ comes close to the \qut{transition} value $N_T$, the curve approaches a plateau and $\mathscr{P}(N)$ becomes independent of $N$. The plateau value in units of the Planck mass in controled by the ratio $L_\mrm{b}/\ell_\mrm{Pl}$, or $N_T$ equivalently:
\bg
\label{eq:Q124}
\lim_{N\to\infty}\mathscr P(N)=(24\pi)^{1/2}\,\frac{m_\mrm{Pl}}{N_T} \, .
\eg
Thus we find that the dimensionful \ac{uv} cutoff $\mathscr P(N)$ never reaches infinity, for no value of $N$ whatsoever.\\ \indent
This result is rather striking. It brings us to our \textbf{main conclusion:}\\

\noindent\textbf{(4a)} In the Background Independent approach the limits $N\to\infty$ and $\mathscr P\to\infty$ are inequivalent. Since, by construction, it is the limit $N\to\infty$ that removes the regularization, it follows that \ita{it is incorrect to attempt taking the limit $\mathscr P\to\infty$ when the gravitational backreaction is taken into account}.\\

\noindent\textbf{(4b)} If, at the classical point $N=0$, the bare radius is of the order of the Planck length or larger, $L_\mrm{b}\gtrsim\ell_\mrm{Pl}$, then no member of the sequence $\{\mathsf{App}(N)\ ,\ 0\leq N<\infty\}$ has a proper \ac{uv} cutoff larger than about the Planck scale: $\mathscr P(N)\lesssim m_\mrm{Pl}$. (We ignore factors of order unity in \gl{Q122} and \gl{Q123}.)\\ \indent
Here we observe a remarkable dynamical mechanism at work which can occur only thanks to the Background Independent quantization scheme: Even though the regulator is removed fully, i.e., all field modes are integrated out, no member $\mathsf{App}(N)$ in the sequence of approximating quantum systems ever encounters transplanckian energies or momenta: The systems dynamically adjust their metric $(g_N^\text{\ac{sc}})\mn$ in such a way that, with respect to this metric, the dimensionful proper cutoff corresponding to the mode cutoff at $n=N$ is always situated below the Planck scale: $\mathscr P(N)\lesssim m_\mrm{Pl}$.

\section{Micro-States of De Sitter Space}
\label{sec:desitter}

The Lorentzian analog of $S^4$, de Sitter space, possesses a Bekenstein-Hawking entropy whose magnitude is determined by the value of the cosmological constant:
\bg
\label{eq:Q130}
\mathscr{S}=\frac{3\pi}{G\Lambda} \, .
\eg
In terms of the Hubble length $L\equiv H^{-1}\equiv (3/\Lambda)^{1/2}$ it reads
\bg
\label{eq:Q131}
\mathscr S=\frac{\pi}{G}L^2
\eg
which is also equivalent to $\mathscr S=\mathscr A/4G$, where $\mathscr A$ denotes the area of the de Sitter horizon, $\mA=4\pi L^2$. Within the Euclidean approach to black holes and similar thermal spacetimes, the  Euclidean 4-sphere plays an important role in the derivation of this result. It serves as a saddle point which enables the semiclassical evaluation of the $g\mn$-integration \cite{Gibbons:1994cg,Gibbons:1977mu,Gibbons:1976ue}.\\ \indent
Being purely thermodynamic in nature, it is a longstanding question what are the microcopic degrees of freedom that get \qut{counted} by this entropy, and which kind of quantum-statistical mechanics might govern a corresponding hidden layer below the (semi-)classical de Sitter spacetime \cite{Kiefer:2012}.\\ \indent 
The sequences of self-gravitating systems which we have constructed suggest a specific answer to this question: If we evaluate the de Sitter entropy for all members of a sequence,
\bg
\label{eq:Q135}
\mathscr{S}(N)\equiv\frac{\pi}{G}\,L^\text{\ac{sc}}(N)^2 \, ,
\eg
it becomes obvious that \ita{the thermodynamic entropy of any member in the $\mathsf{App}(N)$-sequence is determined by precisely the number of degrees of freedom $\mathbf f(N)$ which have given birth to its particular spacetime}.\\ 

To substantiate our claim, we consider the sequence obtained in Subsection \ref{subsec:lambda-null} as an example. We assume $M=0$ here, and so all results are independent of $\xi$. Using the stress tensor of the first type, we found a sequence of systems having $\Lambda_\mrm{b}=0$ and, according to \gl{P70}, self-consistent radii
\bg
\label{eq:Q136}
L^\text{\ac{sc}}(N)^2=\frac{G}{4\pi}\mathbf f(N) \, .
\eg
They amount to the dynamically fixed cosmological constants 
\bg
\label{eq:Q137}
\Lambda^\text{\ac{sc}}(N)\equiv \frac{3}{L^\text{\ac{sc}}(N)^2}=\frac{12\pi}{G\,\mathbf f(N)} \, .
\eg

\noindent\textbf{(1) Quantized radii.} The absolute dimensionful scale in this solution is set by the Planck length $\ell_\mrm{Pl}\equiv G^{1/2}$, which assumes the same value for all members of the sequence. Indeed, their self-consistent radii (Hubble lengths) are quantized in units of $\ell_\mrm{Pl}$:
\bg
\label{eq:Q138}
L^\text{\ac{sc}}(N)=\lef(\frac{\mathbf f(N)}{4\pi}\ri)^{1/2}\ell_\mrm{Pl}=\frac{N^2}{\sqrt{48\pi}}\,\ell_\mrm{Pl}\lef[1+O\lef(\frac{1}{N}\ri)\ri] \, .
\eg
The corresponding spacetimes are well-behaved for any non-zero number $\mathbf f$ of field modes living in the respective universe. For $\mathbf f=0=N$ its metric degenerates, and so we limit ourselves to $N\geq 1$, i.e., to \ita{universes of a genuinly quantum mechanical origin}.\\

\noindent\textbf{(2) Experiment may request $N<\infty$.} As we saw, the sequence \gl{Q136} can be significant for the cosmological constant problem since, for $N\to\infty$, the effective cosmological constant $\Lambda^\text{\ac{sc}}(N)$ approaches zero. Could we also construct a \ac{qft} limit $N\to\infty$ with a non-zero observed cosmological constant in the traditional way by making bare parameters $N$-dependent?\\ \indent
For the specific sequence $\{\mathsf{App}(N)\}$ under consideration the answer is \qut{no}. By virtue of the universal status of $G$ and $\ell_\mrm{Pl}$ in the present case, the formula \gl{Q137} contains no bare parameter that could be given an appropriate $N$-dependence.\\ \indent
Nevertheless, let us hypothesize that \Gl{Q137} represents (the maximum symmetry simplification of) a valid law of Nature, and furthermore that in the real universe cosmologists have measured a nonzero, positive cosmological constant $\Lambda_\mrm{obs}>0$. Then the unavoidable conclusion is that \ita{the physically realized universe carries only a finite number $\mathbf f(N_\mrm{obs})$ of quantum mechanical degrees of freedom rather than a full-fledged quantum field.} The integer $N_\mrm{obs}$ is fixed by the measurement then:
\bg
\label{eq:Q139}
\Lambda^\text{\ac{sc}}(N_\mrm{obs})\overset{!}{=}\Lambda_\mrm{obs} \, .
\eg
In this manner we allow experiment to inform us that \ita{the limit $N\to\infty$ must not be taken in this particular case.}\\ \indent
If instead the measurement reveals that $\Lambda_\mrm{obs}=0$, we do have to let $N\to\infty$, thus activating all field modes. It should be clear though that \qut{$N_\mrm{obs}<\infty$} is a \ita{physical} statement about the matter contents of the universe. Hence the implication that the limit $N\to\infty$ must not be taken has a completely different logical status than the \ita{mathematical inequivalence} of the limits $N\to\infty$ and $\mathscr P\to\infty$ discussed earlier.\\ \indent
In this paper we insisted repeatedly that the approximants $\mathsf{App}(N)$ are more than a generic regularization of a quantum field theory, namely physical systems per se, with finitely many degrees of freedom. One of the virtues of this requirement is that it admits the possibility that \ita{spacetime supports fewer degrees of freedom than would be supplied by a full-fledged quantum field}.\\ \indent
Mindful of this possibility, let us now return to the entropy of de Sitter space.\\

\noindent\textbf{(3) The entropy counts field modes.} Inserting \gl{Q136} into \gl{Q135} we obtain the following semiclassical entropies along the sequence $\{\mathsf{App}(N)\}$:
\bg
\label{eq:Q140}
\mathscr S(N)=\frac{1}{4}\mathbf f(N) \, .
\eg
This is indeed a remarkable result. It confirms what we claimed above: For all members in the chain of systems $\{\mathsf{App}(N)\}$, the Bekenstein-Hawking entropy of their spacetime equals (up to a factor of $1/4$) the number of scalar degrees of freedom that live on this spacetime. Hence the area of the respective Hubble spheres, $\mA(N)$, is an integer multiple of the fundamental unit $\ell_\mrm{Pl}^2$,
\bg
\label{eq:Q141}
\frac{\mA(N)}{\ell_\mrm{Pl}^2}=\mathbf f(N) \, .
\eg\indent
These findings are in line with the intuitive picture that the horizon surface is a fuzzy 2-sphere made of discrete \qut{tiles}. Points on this surface can be distinguished only if their angular separation is larger than about $\Delta\alpha\approx\pi/N$, which implies
\bg
\label{eq:Q145}
\Delta\ell\equiv L^\text{\ac{sc}}(N)\Delta\alpha\approx\sqrt{\pi/48}\,N\,\ell_\mrm{Pl}\lef\{1+O\lef(\frac{1}{N}\ri)\ri\}
\eg
for their approximate proper distance.\\

\noindent\textbf{(4) A \qut{$\Lambda$-$\mN$ connection} proven.} Now let us assume that, as described above, experiment has provided us with some value $\Lambda_\mrm{obs}>0$. Knowing that
\bg
\label{eq:Q150}
\Lambda_\mrm{obs}=\frac{12\pi}{G\,\mathbf f(N)} \, ,
\eg
this piece of data fixes a certain finite $N_\mrm{obs}$, which in turn allows us to infer how many degrees of freedom participated in the, purely quantum dynamical, generation of the universe the measurement of $\Lambda_\mrm{obs}$ was performed in:
\bg
\label{eq:Q151}
\frac{1}{4}\#(\text{degrees of freedom})=\frac{1}{4}\,\mathbf f(N_\mrm{obs})=\frac{3\pi}{G\,\Lambda_\mrm{obs}} \, .
\eg
So we have proven the following property of the de Sitter spaces governed by the $N$-sequence under consideration: Up to the (probably inessential) factor $1/4$, \ita{universes with a strictly positive cosmological constant are described by a quantum theory with only a finite number of degrees of freedom, and this number is given by the Bekenstein-Hawking entropy of de Sitter space}.\\ \indent
This relationship is a particular instance of the conjectured \qut{$\Lambda$-$\mN$-connection} and \qut{$\mN$-bound} that have been speculated about in the literature \cite{Banks:2000fe,Bousso:2000nf} under more general conditions.\footnote{Originally, the $\mN$-bound grew out of string theory based arguments which hinted at the possibility of a $\Lambda$-$\mN$-connection such that all universes with a positive cosmological constant are governed by a fundamental quantum theory with a finite number of degrees of freedom only, whereby this number, $\mN$, is determined by the value of $\Lambda$ \cite{Banks:2000fe,Bousso:2000nf}. For a related discussion within Loop Quantum Gravity we refer to \cite{Smolin:2002sz}. Along an independent line of research, similar indications were found within the Asymptotic Safety approach to Quantum Gravity \cite{Becker:2014pea}. For a counting of modes on dS space related to the emergent gravity paradigm see~\cite{Pad}.} In its stronger form, the hypothesis of the $\mN$-bound claims that in any universe with a positive cosmological constant $\Lambda$, and arbitrary matter contents, the observed entropy $\mathscr S_\mrm{obs}$ is always bounded above: $\mathscr S_\mrm{obs}\leq 3\pi/G\Lambda\equiv\mN$. Obviously our exact result \gl{Q151} precisely matches this claim, and saturates the bound with $\mathscr S_\mrm{obs}=\frac{1}{4}\,\mathbf f(N_\mrm{obs})$ corresponding to the number $\mN$.\\ 

\noindent\textbf{(5) Our universe.} Despite its highly idealized character, it is nevertheless tempting to apply this $N$-sequence to the real universe we live in. If we model the latter by an empty de Sitter space with the observed Hubble radius $L^\text{\ac{sc}}(N_\mrm{obs})\approx10^{60}\ell_\mrm{Pl}$, then \Gl{Q138} yields $N_\mrm{obs}\approx10^{30}$. This implies $\mathscr S(N_\mrm{obs})\approx 10^{120}$ and an angular uncertainty of $\delta\alpha\approx 10^{-30}$. By \gl{Q145} this uncertainty corresponds to a proper length of about $\delta\ell\approx 10^{30}\ell_\mrm{Pl}\approx 10^{-3}\mrm{cm}$ at the present time. Remarkably enough, a degree of fuzzyness of the same order of magnitude has been found in \cite{Reuter:2005bb} by logically independent arguments based upon the functional renormalization group for gravity \cite{Reuter:2019byg}.\\ \indent 
We shall come back to the entropy of de Sitter space elsewhere \cite{N-3} where we also discuss quantum corrections to its Bekenstein-Hawking value.

\section{Summary and Outlook}
\label{sec:conclusions}

In this paper we considered quantum fields in contact with dynamical gravity and proposed a new nonperturbative framework for the efficient investigation of such systems.\\ 

\noindent\textbf{(1)} We advocated the point of view that the principle of Background Independence should apply already to the regularized precursors of a quantum field theory, the \qut{approximants}. To achieve this, we identified three requirements the corresponding calculational scheme must meet. In particular the approximants should constitute \qut{quasi-physical} systems, meaning that, at the very least, they can be ascribed a well defined, finite number of quantum mechanical degrees of freedom which are exposed to, and self-consistently backreact onto the gravitational field.\\ \indent 
In this setting, we reinterpreted the process of removing the regularization as a comparison of such quasi-physical systems. Its limit, if it exists, generalizes the usual continuum limit in that it provides additional information about the spacetime the quantum field prefers to \qut{live in}. The spacetime geometry which is actually realized gets selected in a dynamical fashion, in harmony with Background Independence.\\ \indent
At a more technical level, we introduced what we called cutoffs of the $N$-type. They allowed us to concretely construct sequences of gravity-coupled approximants $\{\mathsf{App}(N)\}$ in terms of the field degrees of freedom. The rationale behind this cutoff scheme is to disentangle the logically unrelated concepts of a regularization parameter on one side, and a proper momentum scale on the other. Besides a careful metric independent enumeration of the field modes, eigenfunctions of the kinetic operator, the installation of an $N$-cutoff requires merely a bisection of the pertinent spectrum, i.e., a rule deciding about to retain, or not to retain a given mode. Since this rule does not involve the metric, we were able to perform limits of such $\{\mathsf{App}(N)\}$-sequences which could not be considered within the standard setting.\\

\enlargethispage{\baselineskip}
\noindent\textbf{(2)} The proposed requirements \textbf{(R1,2,3)} for an efficient quantization scheme were outlined and motivated in Section \ref{sec:framework}. Their interpretation is as follows: At a given level of technical complexity, it is the calculational schemes which do meet the requirements that have the best chances to get close to the true physical answers. In a way, the motivation for using gravitationally backreacting approximants is similar to the (well-justified) hope that theories with a symmetry are best analyzed in a regularization scheme that respects this symmetry.\\ \indent 
The proviso of a fixed level of complexity may be important here. While there are good reasons to believe that \textbf{(R1,2,3)} can help us finding the fastest track to the correct answers, this does not exclude the possibility that the same answers can also be found by other means, albeit at a higher price.\\

\noindent\textbf{(3)} As a first example, we applied the set of rules \textbf{(R1,2,3)} to a Gaussian scalar field. It was quantized in a universe on whose metric it was allowed to backreact selfconsistenly. We focussed on maximally symmetric spacetimes, and this enabled us to perform all subsequent calculations \ita{exactly}. In particular we never had to invoke the short time expansion of heat kernel traces; while this is a frequently used tool in such studies, usually it has the disadvantage of generating asymptotic series only. The exact calculations are a further source of differences between the present approach and earlier investigations.\\ \indent
The operator which represents the scalar's energy-momentum tensor at the quantum level is not unique. In the standard approach, products of field operators at coincident points must be given a meaning. This leads to ambiguities and requires external input over and above the classical theory. The same is true in the present approach. We employed two types of quantum stress tensors, which turned out inequivalent as for their detailed predictions.\\ \indent
And yet, using either of the stress tensors we found sequences of approximants which displayed the same astounding phenomenon: \ita{adding further degrees of freedom to the quantum system flattens the universe}. This is exactly the opposite of what background dependent calculations like Pauli's estimate predict, namely that further modes imply higher curvature.\\ \indent 
We demonstrated explicitly that, as shown in Figure \ref{fig:D}, \ita{taking the \ac{qft} limit of infinitely many field modes, and including the quantum system's backreaction on the metric, are two non-commuting operations}.\\ \indent
In view of these findings we believe that contrary to many claims there does not exist any \qut{cosmological constant problem} due to quantum vacuum fluctuations. The false impression of a huge induced cosmological constant arises only if one takes the wrong path in the diagram of Figure \ref{fig:D}, namely the one which relies on the dangerous illusion that there can be a rigid spacetime that would never change, whatever load of energy and momentum is imposed on it.\\

\noindent\textbf{(4)} Our final application was to the thermodynamics of de Sitter space. We could prove a \qut{$\Lambda$-$\mN$ connection} of the kind speculated about in the literature on entirely different grounds. Computing the semiclassical Bekenstein-Hawking entropy for the approximants along one of our sequences, we found that a de Sitter spacetime with a given cosmological constant is to be identified with \ita{one specific approximant at finite $N$}, rather than a limit of such. Remarkably enough, its Bekenstein-Hawking entropy turned out to count the number of field modes that are governed by this approximant, thus suggesting a natural interpretation for the micro-states of de Sitter spacetime.\\ 

\noindent\textbf{(5)} It is clear that the present work should be extended in a number of directions. In this paper the discussion mostly focused on the approximants per se. Future work will have to find more physically realistic approximants, and to determine which sequences actually converge to quantum field theories with desirable properties. This will require including self-interactions in the matter sector \cite{N-3}. The next steps also include an application to full Quantum Gravity, see \cite{N-2} for a first investigation. 

\vspace{0.2cm}
\noindent
\subsubsection*{Acknowledgments}
This work is supported by DFG Grant No. RE 793/8-1.

\clearpage


\end{document}